\documentclass[12pt]{article}
\usepackage{algorithm2e}
\usepackage{amscd}
\usepackage{amsfonts}
\usepackage{amsmath}
\usepackage{amssymb}
\usepackage{amsthm}
\usepackage{authblk}
\usepackage[english]{babel}
\usepackage{bm}
\usepackage{caption}
\usepackage{color}
\usepackage{float}
\usepackage[perpage, symbol*]{footmisc}
\usepackage{graphicx}
\usepackage{hyperref}
\usepackage{enumerate}
\usepackage{mathrsfs}
\usepackage{natbib}
\usepackage{setspace}
\usepackage{subcaption}
\usepackage{tikz}
\usepackage{url}

\newcommand{\blind}{1}

\DeclareMathOperator*{\argmin}{arg\,min}
\newcommand{\T}{\intercal}

\theoremstyle{definition}
\newtheorem{definition}{Definition}

\setcitestyle{aysep={}}

\begin{document}

\def\spacingset#1{\renewcommand{\baselinestretch}%
{#1}\small\normalsize} \spacingset{1}


\if1\blind
{
  \title{\bf Longitudinal Principal Manifold Estimation}
  \author{
    Robert Zielinski\hspace{.2cm}\\
    Department of Biostatistics, Brown University\\
    Kun Meng \\
    Division of Applied Mathematics, Brown University\\
    and \\
    Ani Eloyan \\
    Department of Biostatistics, Brown University\\
    for the Alzheimer's Disease Neuroimaging Initiative\thanks{Data used in preparation of this article were obtained from the Alzheimer's Disease Neuroimaging Initiative (ADNI) database (adni.loni.usc.edu). As such, the investigators within the ADNI contributed to the design and implementation of ADNI and/or provided data but did not participate in analysis or writing of this report. A complete listing of ADNI investigators can be found at \href{http://adni.loni.usc.edu/wp-content/uploads/how_to_apply/ADNI_Acknowledgement_List.pdf}{\texttt{http://adni.loni.usc.edu/wp-content/uploads/how\_to\_apply/ADNI\_Acknowledgement\_List.pdf}}}}
  \maketitle
} \fi

\if0\blind
{
  \bigskip
  \bigskip
  \bigskip
  \begin{center}
    {\LARGE\bf Longitudinal Principal Manifold Estimation}
\end{center}
  \medskip
} \fi

\bigskip
\begin{abstract}
Longitudinal magnetic resonance imaging data is used to model trajectories of change in brain regions of interest to identify areas susceptible to atrophy in those with neurodegenerative conditions like Alzheimer’s disease. Most methods for extracting brain regions are applied to scans from study participants independently, resulting in wide variability in shape and volume estimates of these regions over time in longitudinal studies. To address this problem, we propose a longitudinal principal manifold estimation method, which seeks to recover smooth, longitudinally meaningful manifold estimates of shapes over time. The proposed approach uses a smoothing spline to smooth over the coefficients of principal manifold embedding functions estimated at each time point. This mitigates the effects of random disturbances to the manifold between time points. Additionally, we propose a novel data augmentation approach to enable principal manifold estimation on self-intersecting manifolds. Simulation studies demonstrate performance improvements over naïve applications of principal manifold estimation and principal curve/surface methods. The proposed method improves the estimation of surfaces of hippocampuses and thalamuses using data from participants of the Alzheimer’s Disease Neuroimaging Initiative. An analysis of magnetic resonance imaging data from 236 individuals shows the advantages of our proposed methods that leverage regional longitudinal trends for segmentation.
\end{abstract}

\noindent%
{\it Keywords:}  Alzheimer's disease, smoothing spline, magnetic resonance imaging
\vfill

\newpage
\spacingset{1.9}
\section{Introduction}

Neuroimaging plays a critical role in the diagnosis and monitoring of a number of common neurodegenerative conditions, such as Alzheimer's disease (AD) \citep{knopmanAlzheimerDisease2021}. Frequently, interest centers around longitudinal changes in one or more neurological substructures. These structural changes can be observed in magnetic resonance imaging (MRI) data \citep{crainiceanu2016tutorial}. For example, it is common to observe atrophy in the hippocampus, a structure in the temporal lobe of the brain, of those with  cognitive impairment either due to AD or other causes. Image segmentation enables the extraction of subcortical structures from MRI images of the brain for identifying differences between these structures in disease populations, modeling trajectories of change in the structures, and evaluating treatment effects in terms of reduction of atrophy over time. Traditionally, manual segmentation of images by a trained radiologist has been considered the most accurate approach for segmentation of regions of interest and is the gold standard. However, this approach is highly time and resource intensive. When analyzing data from studies with a large number of images, this approach may impose prohibitive costs. Automated segmentation approaches, including FSL-FIRST, have been introduced to address these concerns \citep{patenaudeBayesianModelShape2011}. While automating the segmentation process drastically reduces costs, it potentially introduces additional inaccuracies.

Failures of automatic segmentation algorithms may have dramatic effects when analyzing large datasets of images where manual quality control may not be feasible, such as data from the Alzheimer's Disease Neuroimaging Initiative (ADNI) study. In a study comparing the accuracy of FIRST and FreeSurfer, another automated segmentation method, using a large sample, \cite{mulderHippocampalVolumeChange2014} found that 6.9\% of segmentations by FIRST failed visual inspection for accuracy, as did 7.5\% of segmentations by FreeSurfer. After removing the scans with failed segmentations, FIRST and FreeSurfer produced segmentations with variability similar to and slightly lower than manual segmentation, respectively. If failed segmentations were not removed from analysis, reflecting a more realistic situation when working with a large number of images, variability was much higher for FIRST and FreeSurfer than for manual segmentation. Ultimately, high levels of variability are present at the subcortical level, both between study visits for the same individual and between individuals, regardless of the segmentation approach, and may be particularly influential when using automated segmentation methods. Mitigating the extent of this variability is a priority from a statistical perspective to improve the inferences that can be made from segmented image data. To this end, in this article, we propose a manifold learning-based method to develop smooth estimates of the surfaces of subcortical structures over time.

Manifold learning refers to a set of approaches to modeling high-dimensional data that satisfy the ``manifold hypothesis"---\textit{``high-dimensional data tend to lie in the vicinity of a low-dimensional manifold''} \citep{fefferman2016testing}. Manifold learning approaches have been applied in various scientific fields, e.g., single-cell biology \citep{ding2023learning} and robotics \citep{gao2023k}. In medical imaging applications, interacting with a low-dimensional interpretation of a structure may be more intuitive and efficient than working with the structure in its original high-dimensional space. For example, \cite{yueParameterizationWhiteMatter2016} seeks to compute a parametric representation of the corpus callosum that is used for investigating the features of this structure in people with multiple sclerosis. Estimating time-dependent low-dimensional representations of these shapes is essential for investigating changes of these low-dimensional structures. Nonlinear manifold learning methods have been extensively studied in the literature. However, many popular approaches, such as Isomap \citep{tenenbaumGlobalGeometricFramework2000}, locally linear embedding \citep{roweisNonlinearDimensionalityReduction2000}, and Laplacian eigenmaps \citep{belkin2003laplacian}, are not applicable in longitudinal studies. These methods include the time dimension in the dimension reduction process, causing the time dimension to be incorporated into the low-dimensional parameterization and preventing meaningful interpretation of the time dimension.

Previous work on longitudinal dimension reduction methods has primarily focused on linear approaches, such as the longitudinal functional PCA method proposed by \cite{greven2011longitudinal}. As for nonlinear methods, \cite{wolzManifoldLearningBiomarker2010} used Laplacian eigenmaps to incorporate longitudinal information when reducing neuroimaging data to lower dimensional space. However, parameterizations for baseline images and longitudinal image differences were estimated separately while we seek to parameterize them simultaneously. Additionally, \cite{louisRiemannianGeometryLearning2019} used a deep learning-based approach to learn Riemannian manifolds in the context of disease progression modeling. However, this approach relies on modeling assumptions specific to the application, and the deep learning methodology used imposes costs in terms of computational requirements and interpretability. Finally, \cite{busch2023Multiview} developed the T-PHATE manifold learning algorithm to account for the autocorrelation present in functional MRI (fMRI) data, but the temporal modeling also depends on assumptions specific to the properties of fMRI data.

In this article, we propose a new longitudinal approach to estimating nonlinear manifolds over time. Rather than treating the time dimension the same as the space dimensions, we introduce a novel process that maintains the interpretability of the time dimension. Specifically, we use the method proposed by \cite{mengPrincipalManifoldEstimation2021} to model the appropriate principal manifold at each given time point. Unlike most popular manifold learning methods, this approach results in an analytic and computationally efficient functional representation of the manifold estimate (see equation \eqref{eq: spline representation of the PME}). We build on this model by imposing smoothness over these approximated manifolds in the time dimension, yielding an estimate showing longitudinal changes in the underlying manifold. We demonstrate that the regularization involved in this process mitigates the effects of variability between time points. Our work includes two major contributions to the existing literature: 1) we propose a novel approach for longitudinal smoothing in the general-purpose manifold learning setting, and 2) we show the potential of leveraging longitudinal manifold smoothing for reducing variability and increasing signal-to-noise in segmentation of brain structures over time.

This work was motivated by longitudinal imaging studies, where data of participants are collected during several study visits to observe changes in brain structure (among other variables of interest) over time. One such study is the ADNI, where we obtained data that we analyzed to reach the results presented in this article. The data are publicly available and are hosted at \href{https://adni.loni.usc.edu/}{\texttt{https://adni.loni.usc.edu/}}. The study has over 20 years of history with the goal of obtaining biomarkers of progression of early AD from various data sources, including brain imaging (such as magnetic resonance imaging and positron emission tomography), cognitive and biological markers. We implement the proposed methods to estimate the surfaces of two regions of interest in the brain - the hippocampus and the thalamus. The thalamus is selected as a ``easy-to-fit" region of interest to evaluate the performance of our proposed methods when the underlying structure is somewhat close to a spherical shape. The hippocampus is more of a pear-shaped region and is selected as a region that may exhibit atrophy for people with AD.

The remainder of this article is laid out as follows. Section \ref{s:PME} reviews the principal manifold framework proposed by \cite{mengPrincipalManifoldEstimation2021}. In Section \ref{s:LPME}, we adapt the principal manifold framework for longitudinal settings, and show our proposed novel algorithm for dimension reduction using the longitudinal principal manifold framework. Section \ref{s:simulations} demonstrates the performance of this approach on simulated data. In Section \ref{s:application}, we apply our proposed method to estimate smooth time dependent surfaces of the hippocampus and thalamus in ADNI participants. The article concludes with a discussion of the method's contributions in Section \ref{s:discussion}.

\section{Principal Manifold Estimation Algorithm}\label{s:PME}

The framework for principal manifolds originated with the concept of principal curves \citep{hastiePrincipalCurves1989}, which are essentially curves that pass through the middle of a data cloud. Motivated by the penalization approach presented by \cite{smolaRegularizedPrincipalManifolds2001}, \cite{mengPrincipalManifoldEstimation2021} proposed the principal manifold estimation (PME) framework, which extends principal curves to arbitrarily high intrinsic dimensions. In order to present the longitudinal PME approach, we start by introducing notation used throughout this paper. We use $d$ and $D$ to denote the dimensions of the low-dimensional manifold and high-dimensional spaces, respectively. We assume that a data cloud is observed in the $D$ dimensional space. In the context of segmentation, we consider a collection of vertices on the surface of the brain region of interest as the 3-dimensional data cloud ($D=3$), while the 2-dimensional smooth surface of the region is the underlying manifold of interest ($d=2$; for example, see Figure S1 in the online supplement). For any positive integer $q$ and point $\xi=(\xi_1,\ldots,\xi_q)\in\mathbb{R}^q$, we use the norm $\Vert \xi\Vert_{\mathbb{R}^q}=\sqrt{\sum_{l=1}^q \xi_l^2}$. Since our interest in this work centers on continuous mappings, we denote the collection of continuous vector-valued functions from $\mathbb{R}^d$ to $\mathbb{R}^D$ as $\mathcal{C}(\mathbb{R}^{d} \to \mathbb{R}^{D})$. Given a $d$-variable function $u$, $\nabla^{\otimes 2} u$ denotes the Hessian matrix of $u$, defined as $r \to \left( \frac{\partial ^2 u}{\partial r_i \partial r_j}(r) \right)_{1 \leq i, j \leq d}$. The $d$-dimensional manifold corresponding to a function $f \in C(\mathbb{R}^{d} \to \mathbb{R}^{D})$ is formally defined as $M_f^d=\{f(r):\,r\in\mathbb{R}^d\}$. Many manifold estimation methods are based on defining a projection index, denoted by $\pi_f(x)$, defined as the point in the $d$-dimensional space, where $f(\pi_f(x))$ is the projection of $x$ on the manifold $M_f^d$. The definitions of $\pi_f(x)$ given by \cite{hastiePrincipalCurves1989} and \cite{mengPrincipalManifoldEstimation2021} for one- and higher-dimensional intrinsic spaces, respectively, guarantee the uniqueness of the projection. The principal manifolds are defined as follows.
\begin{definition}
  \label{def:principal_manifolds}
  Let $\mathbf{X}$ denote a random $D$-vector associated with the probability distribution $\mathbb{P}$ such that $\mathbf{X}$ has compact support, $\text{supp}(\mathbb{P})$, and finite second moments. Under some assumptions on limiting behavior and smoothness on function $f$, given a $\lambda \in [0, \infty)$, define the following functional:
  \begin{align}\label{eq:pme_kappa}
\mathcal{K}_{\lambda, \mathbb{P}}(f) = \mathbb{E}\|\mathbf{X} - f(\pi_f(\mathbf{X}))\|_{\mathbb{R}^{D}}^2 + \lambda\|\nabla^{\otimes 2}f\|_{L^2(\mathbb{R}^{d})}^2.
  \end{align}
  The principal manifold of $\mathbf{X}$ with the tuning parameter $\lambda$ is the manifold $M_{f^{*}}^{d}$ determined by $f^{*}$ if
  $f_{\lambda}^{*} = \argmin_{f \in \mathcal{F}(\mathbb{P})}\mathcal{K}_{\lambda, \mathbb{P}}(f)$,
where $\mathcal{F}(\mathbb{P})$ is the collection of functions $f \in \mathcal{C}(\mathbb{R}^{d} \to \mathbb{R}^{D})$ such that $\lim_{\|r\|_{\mathbb{R}^{d}} \to \infty}\|f(r)\|_{\mathbb{R}^{D}} = \infty$, the components of the function $f$ are in $\nabla^{-\otimes 2}L^2(\mathbb{R}^{d}) = \left\{u : \|\nabla^{\otimes 2} u\|_{\mathbb{R}^{d \times d}} \in L^2(\mathbb{R}^{d})\right\}$, and the projection function of $f$ satisfies $\sup_{x \in \text{supp}(\mathbb{P})}\|\pi_f(x)\|_{\mathbb{R}^{d}} = 1$.
\end{definition}

It is important to note that, unlike manifold learning methods that use an eigendecomposition to directly estimate coordinates in the $d$-dimensional manifold space with only an implicit representation of the embedding function $f$ (e.g. Isomap, locally linear embedding, and others), principal manifolds may be estimated via an explicit approximation of the embedding function $f$, from which the projection index $\pi_f(x)$ and the desired coordinates on the manifold can be found. In this way, regression approaches become relevant to the manifold estimation problem.

The functional $\mathcal{K}_{\lambda, \mathbb{P}}(f)$ in equation \eqref{eq:pme_kappa} consists of two elements. The first element is the expected squared distance from data points $\mathbf{X}$ in the cloud to their projections $f(\pi_f(\mathbf{X}))$ on the manifold $M_f^d$. The second element $\|\nabla^{\otimes 2}f\|_{L^2(\mathbb{R}^{d})}^2 = \sum_{l=1}^{D} \int_{\mathbb{R}^{d}}\sum_{i, j = 1}^{d}\left|\frac{\partial^2f_l}{\partial r_i \partial r_j}(r)\right|^2dr$ imposes the smoothness/curvature penalty on the estimated manifold. The coefficient $\lambda$ is a tuning parameter. In this setting,  $\pi_{f_{\lambda}^{*}}(\mathbf{X})$ maps the $D$-vector $\mathbf{X}$ to a $d$-dimensional parameterization, while $f_{\lambda}^{*}$ embeds the $d$-dimensional parameterization in the original $D$-dimensional space. Thus, even though manifold learning is commonly considered an unsupervised learning approach, \cite{mengPrincipalManifoldEstimation2021} show that finding the function that minimizes $\mathcal{K}_{\lambda, \mathbb{P}}(f)$ in fact takes the form of a penalized regression problem, making this a supervised learning problem. \cite{mengPrincipalManifoldEstimation2021} also show that this function takes spline form
\begin{align}\label{eq: spline representation of the PME}
    f_{(n+1), l}(r) = \sum_{j=1}^N s_{j, l} \times \eta_{4-d}\left(r - \pi_{f_{(n)}}(\mu_{j, N})\right) + \sum_{k=1}^{d + 1}\alpha_{k, l} \times p_k(r), \ \ \text{ for } \ l = 1, 2, \dots, D,
\end{align}
under the constraint $\sum_{j=1}^{N}s_{j, l} \times p_{k}\left(\pi_{f_{(n)}}(\mu_{j, N})\right) = 0$ for all $k = 1, 2, \dots, d + 1$ and $l = 1, 2, \dots, D$, where $\mu_{j, N}, j = 1, \dots, N$ denote points that summarize the data cloud. Importantly, $N$ tends to be much smaller than the number of points in the data cloud \citep[fig. 3]{mengPrincipalManifoldEstimation2021}, which indicates that the representation in equation \eqref{eq: spline representation of the PME} is computationally efficient. This spline function is specified by coefficients $s_{j, l}$ and $\alpha_{k, l}$ for $j = 1, \dots, N$, $k = 1, \dots, d+1$, and $l = 1, \dots, D$. These coefficients will be used to smooth over principal manifold estimates obtained at several time points in Section \ref{s:LPME}.

\section{Longitudinal Principal Manifold Estimation}\label{s:LPME}

In this section, we introduce an approach to extending the PME framework described in Section \ref{s:PME} to longitudinal point clouds. We first heuristically explain the goal of our proposed model and its importance in the segmentation of brain regions for longitudinal imaging studies. Our goal is to estimate a smooth surface for the brain region of interest at each time point such that the changes of the surface over time are smooth. The estimated surfaces across different time points enable further inference, e.g., modeling trajectories of change in region volumes over time, identifying regions that are most affected in terms of atrophy, etc. We consider that 3-dimensional data clouds are observed longitudinally during several visits by a study participant. Hence, the observed data $(x_1, x_2, x_3,t_j)$ are 4-dimensional, where $t_j$ denotes the time of the $j$th study visit, and $(x_1, x_2, x_3)$ denotes a spatial point on the surface of the segmented brain region of interest.

Figure \ref{fig:lpme_step1} displays a simplified example of this setting using simulated data, in which 2-dimensional point clouds are observed longitudinally at several time points. The goal in this simplified example is to fit a surface in 3-dimensional spacetime---comprising two spatial dimensions and one time dimension---that fits the longitudinal data $(x_1, x_2, t_j)$ in spacetime and is not affected by the random noise added at each time point and in space (see Figure \ref{fig:lpme_step4}). The PME framework does not directly apply due to the following: (1) Suppose we apply the PME approach to the data $(x_1, x_2, t_j)$ in spacetime and obtain an estimated surface $(r_1, r_2) \mapsto F(r_1, r_2)\in\mathbb{R}^2\times\mathbb{R}$ in spacetime. It is likely that neither coordinate $r_1$ nor $r_2$ coincides with the time dimension---they are (potentially nonlinear) combinations of both time and spatial coordinates, which prevents meaningful interpretation of the time dimension. (2) If the manifold is estimated separately at each time point, the results may not be smooth over time. We expect that our proposed dimension reduction method will yield a smooth surface $(t,r)\mapsto F(t,r)$ in spacetime, $\mathbb{R}^2\times\mathbb{R}$, while ensuring that one coordinate, $t$, coincides with the time dimension. This can be achieved by the longitudinal principal manifold estimation (LPME) framework proposed in this section.

\subsection{The Longitudinal Principal Manifold Framework}\label{section: The Longitudinal Principal Manifold Framework}

Hereafter, we utilize the following notation to describe the LPME framework for generic low and high dimensions, $d$ and $D$. Let $\left\{x_{it}\right\}_{i=1, t=1}^{I_t, T}$ represent the $I = \sum_{t=1}^{T}I_t$ observations in $\mathbb{R}^D$ (i.e., each $x_{it}\in\mathbb{R}^D$) for each image $i$, where $I_t$ denotes the number of observations available at each time point $t$, with $T$ total time points. As in the setting described above, at each time point $t$, these observations lie in the vicinity of a $d$-dimensional manifold and are corrupted by $D$-dimensional noise. In this setting, the manifold to estimate is represented by the following, where $t$ coincides with the time dimension
\begin{align}\label{eq: F form}
    \begin{aligned}
        F:\ \ & \mathbb{R}^d\times\mathbb{R} \rightarrow \mathbb{R}^D\times\mathbb{R},\ \ \  (t,\mathbf{r})\mapsto F(t,\mathbf{r})=:f_t(\mathbf{r}).
    \end{aligned}
\end{align}
We define a longitudinal principal manifold as follows.
\begin{definition}
  \label{def:lpme} Given a collection $\mathbf{X} = \{ \mathbf{X}_t \}_{t=1}^T$ of data clouds observed at a series of time points, where $\mathbf{X}_t$ is the random  $D$-vector observed at time $t$, and tuning parameters $\lambda=\{\lambda_t\}_{t\in\mathbb{R}}$ and $\gamma$, we define the functional $\mathcal{K}_{\lambda, \gamma, \mathbb{P}}(F)$ as follows
\begin{align}\label{eq:newKappa}
      \notag\mathcal{K}_{\lambda, \gamma, \mathbb{P}}(F) &:= \int_\mathbb{R} \mathbb{E}\left\|\mathbf{X}_t - f_t\left(\pi_{f_t}(\mathbf{X}_t)\right)\right\|_{\mathbb{R}^{D}}^2 \, dt + \int_\mathbb{R} \lambda_t \cdot\|\nabla^{\otimes 2}f_t\|_{L^2(\mathbb{R}^{d})}^2 \, dt + \gamma\cdot \int_{\mathbb{R}}\left\|\frac{\partial^2}{\partial t^2}F\right\|_{L^2(\mathbb{R}^d)}^2 \, dt \\
  &= \int_{\mathbb{R}}\mathcal{K}_{\lambda_t, \mathbb{P}}(f_t) \, dt + \gamma \cdot \int_{\mathbb{R}}\left\|\frac{\partial^2}{\partial t^2}F\right\|_{L^2(\mathbb{R}^d)}^2 \, dt,
\end{align}
\sloppy where $f_t(\mathbf{r})=F(t,\mathbf{r})$ is a continuous function of the form in equation \eqref{eq: F form}, satisfying $\lim_{\|\mathbf{r}\|_{\mathbb{R}^{d}} \to \infty, t \to \infty}\|F(t,\mathbf{r})\|_{\mathbb{R}^{D}\times\mathbb{R}} = \infty$, and its coordinates are in $\nabla^{-\otimes 2}L^2(\mathbb{R}^{d}\times\mathbb{R}) = \left\{u: \|\nabla^{\otimes 2} u\|_{\mathbb{R}^{d \times d}} \in L^2(\mathbb{R}^{d}\times\mathbb{R})\right\}$. Then, the manifold $M_{F^{*}}^{d+1}:=\{F^*(t,\mathbf{r}): (t,\mathbf{r})\in \mathbb{R}^d\times\mathbb{R}\}$ is the longitudinal principal manifold of $\mathbf{X}$ if $F_{\lambda}^{*} = \argmin_{F}\mathcal{K}_{\lambda, \gamma, \mathbb{P}}(F)$.
\end{definition}

By minimizing the functional $\mathcal{K}_{\lambda, \gamma, \mathbb{P}}$ in (\ref{eq:newKappa}), our goal is to (i) minimize the distance between each data point $\mathbf{X}_t$ and its projection $f_t\left(\pi_{f_t}(\mathbf{X}_t)\right)$ on the fitted manifold at each time point $t$, (ii) penalize the roughness of the fitted manifold at each time point $t$ depending on the value of the tuning parameter $\lambda_t$, and (iii) impose smoothness of the function $F$ over time regularized by the tuning parameter $\gamma$. These goals correspond to minimizing the first, second, and third additives of the functional $\mathcal{K}_{\lambda, \gamma, \mathbb{P}}$ in (\ref{eq:newKappa}), respectively. Using separate smoothing parameters provides additional flexibility in situations where a manifold with a high level of spatial roughness shows minimal changes over time, or vice versa.

\subsection{The LPME Algorithm}

\begin{figure}
  \centering
  \subfloat[\centering Data]{\label{fig:lpme_step1} \includegraphics[width=7cm]{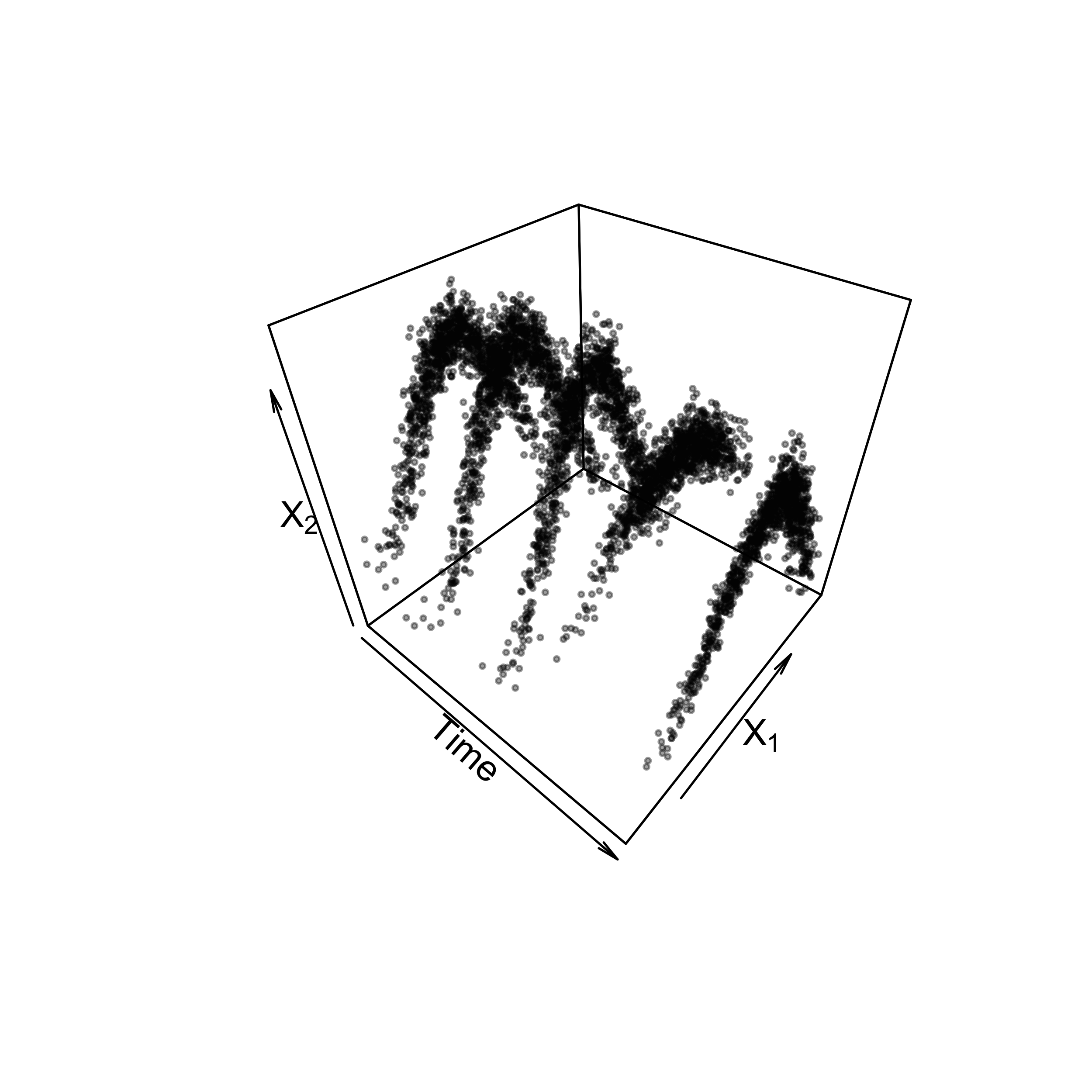}}%
  \hfill
  \subfloat[\centering Step 1: Sample Size Reduction]{\label{fig:lpme_step2} \includegraphics[width=7cm]{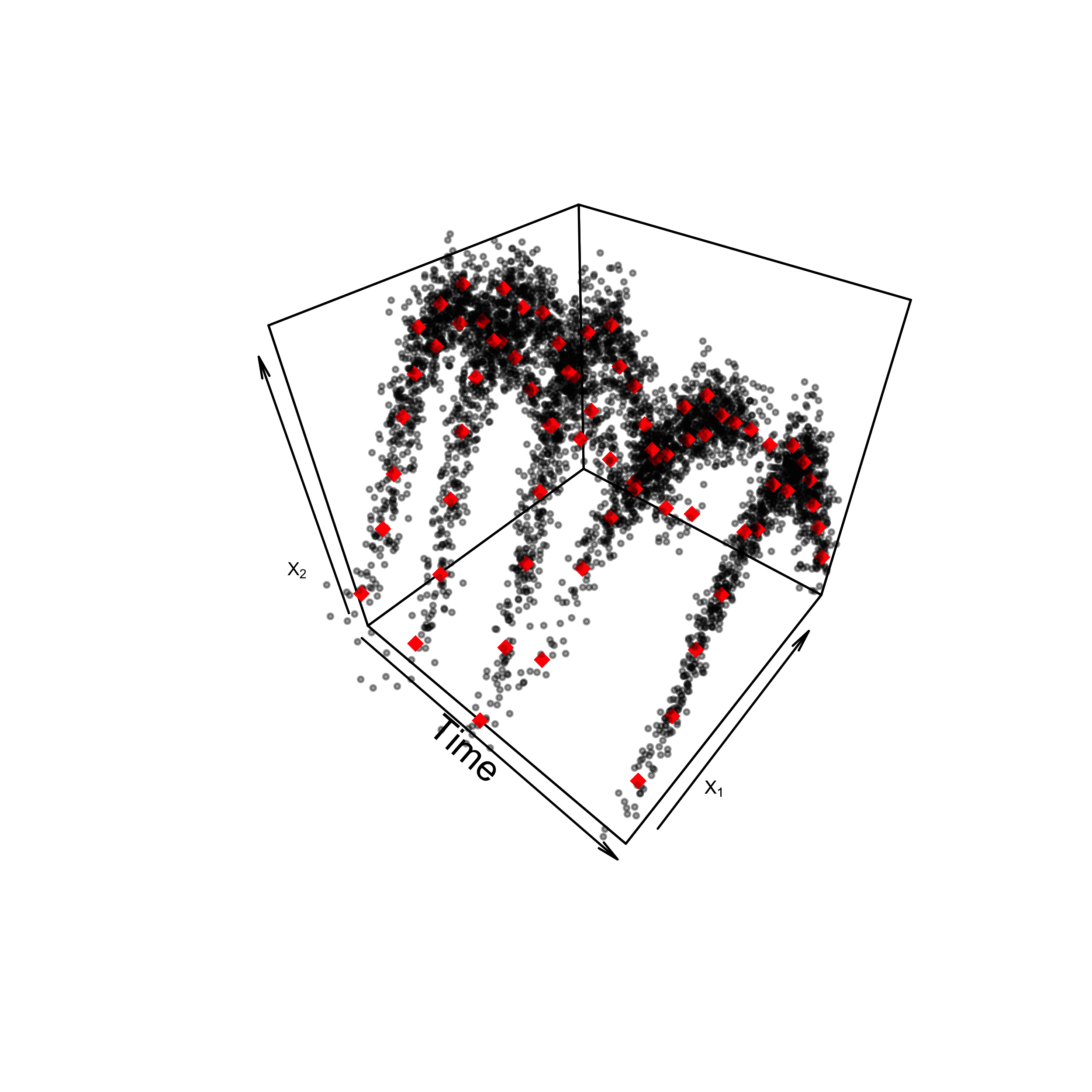}}
  \vfill
  \subfloat[\centering Step 2: Initialization]{\label{fig:lpme_step3} \includegraphics[width=7cm]{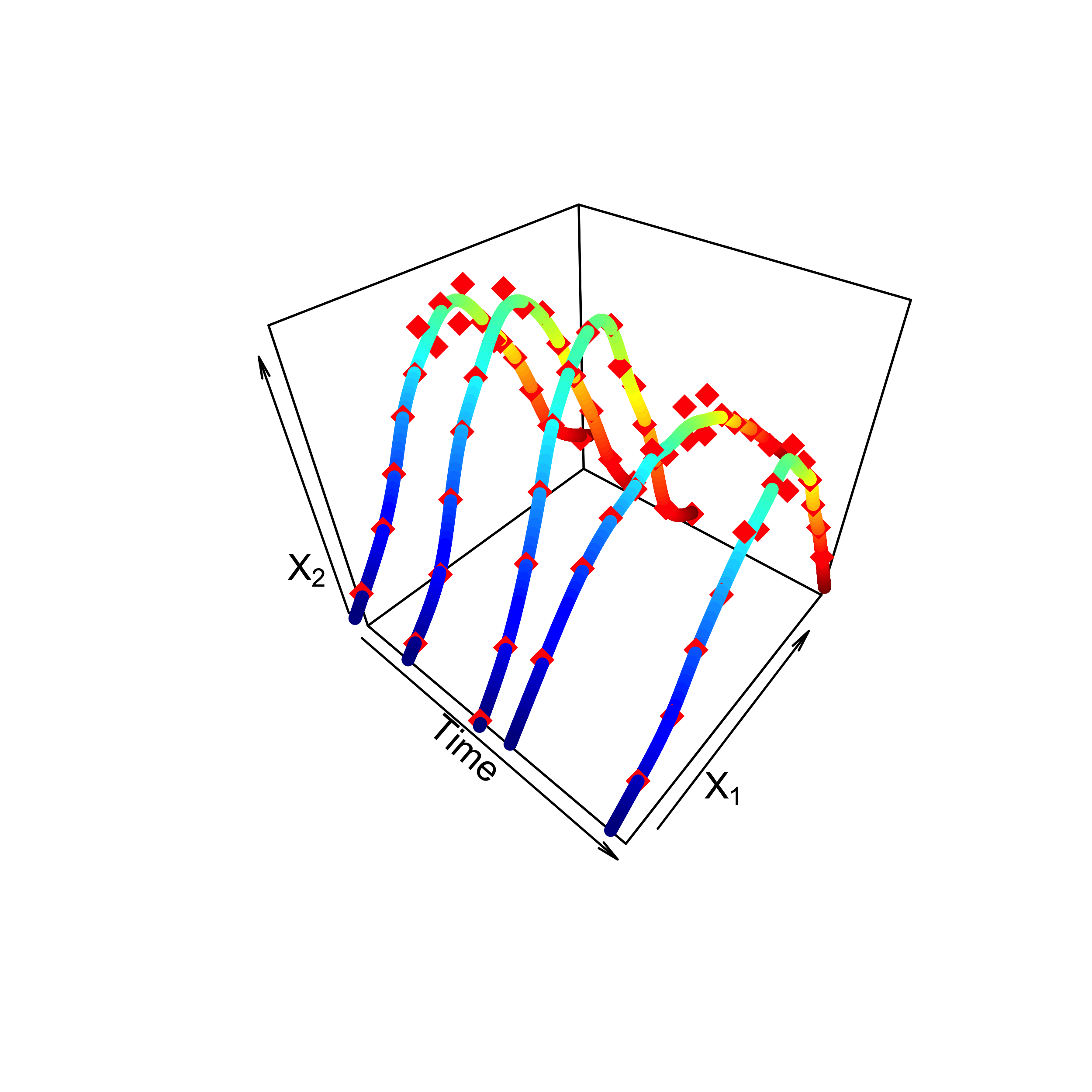}}
  \hfill
  \subfloat[\centering Steps 3 and 4: Fitting and Tuning]{\label{fig:lpme_step4} \includegraphics[width=7cm]{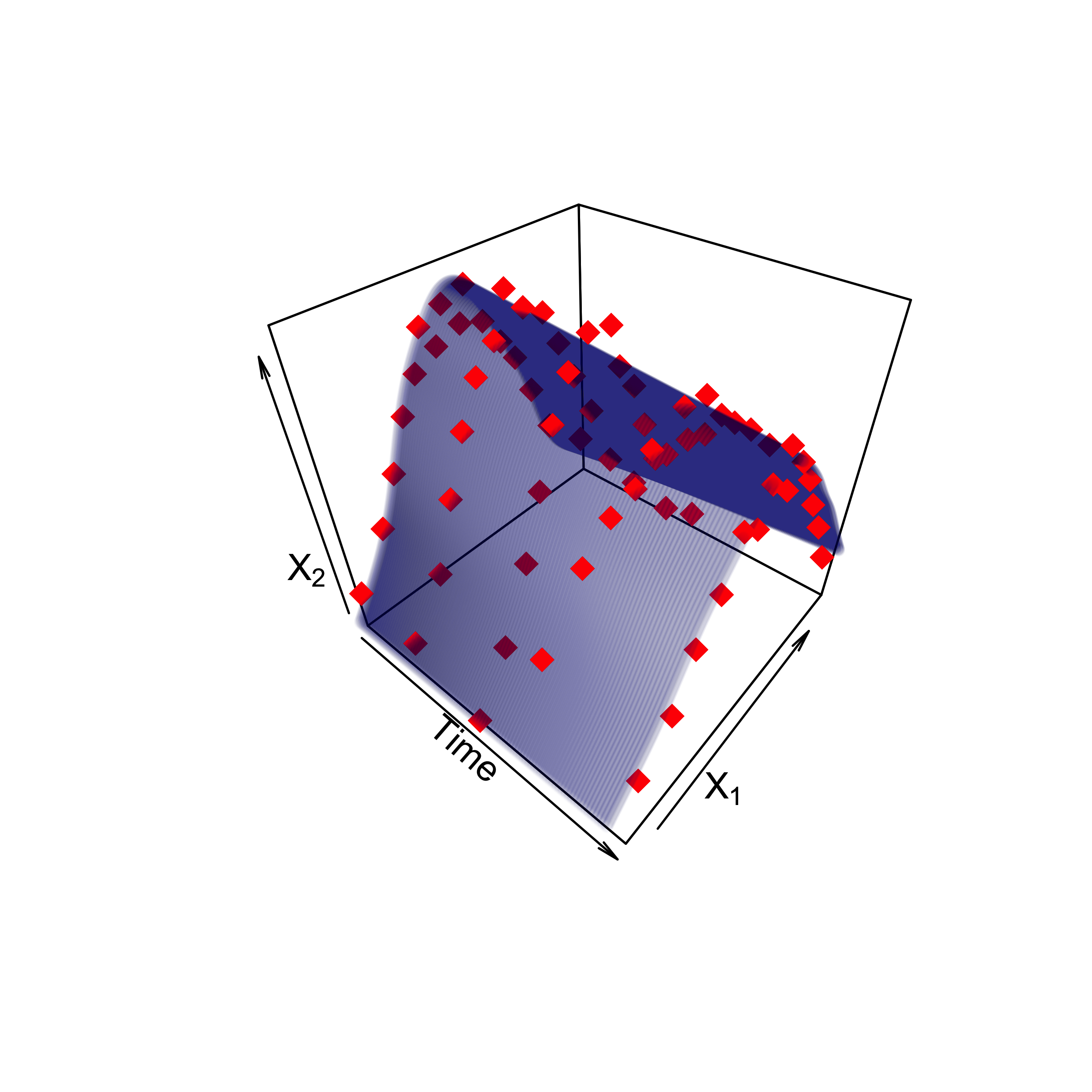}}
  \caption{{\footnotesize LPME algorithm steps, demonstrated using longitudinal simulated data with $D=2$. Data from the functions $x_1 = r$ and $x_2 = \alpha \sin \left(\beta r + \frac{\pi}{2}\right)$ are generated at 5 time points. Random noise is added both in time and space as $\zeta \bf{g}(t) + \iota$, where $\bf{g}(t)$ represents a normally distributed spatial translation applied to all points at each time, $\zeta$ represents a normally distributed amplitude multiplier for this translation, and $\iota$ represents normally distributed noise applied to each point individually. In (a) we show the original data cloud, in (b) we show the sample size reduction step, where the diamond-shaped points indicate the representative data to be used for the next steps of the algorithm, (c) shows the fitted time specific curves, the surface in (d) shows the estimated smooth surface.} }
  \label{fig:lpme_steps}
\end{figure}

Given the framework proposed in Section \ref{section: The Longitudinal Principal Manifold Framework}, estimation of the longitudinal principal manifold entails minimizing the functional in (\ref{eq:newKappa}). Our proposed algorithm is based on a multi-stage approach, where we consider smoothing of the data clouds at each time point (spatial smoothing) and then smoothing of the obtained manifolds over time (temporal smoothing). Specifically, at each individual time point, we apply the PME algorithm \citep[alg. 2]{mengPrincipalManifoldEstimation2021} to fit the data cloud at the time point and represent the fitted manifold using the spline form in equation~\eqref{eq: spline representation of the PME}. Hence, we have a collection of spline coefficients, i.e., the $s_{j,l}$ and $\alpha_{k,l}$ in equation~\eqref{eq: spline representation of the PME}, associated with each time point. Then, we smooth the time-dependent spline coefficients with respect to time. Details of this procedure are encapsulated in our proposed LPME algorithm, which consists of four steps: data reduction, initialization, fitting, and tuning. A visualization of these steps is shown in Figure \ref{fig:lpme_steps}. Details of this approach are described below and formally given in Algorithm \ref{alg:lpme}\if1\blind{, and an implementation is available as an \texttt{R} package at \href{https://github.com/rjzielinski/pme}{\texttt{https://github.com/rjzielinski/pme}}}\fi.

\subsubsection{Data Reduction}

In most practical manifold learning tasks, data are often quite large. We begin by deriving a small number of feature points in $D$-dimensional space that capture the intrinsic structure of the data cloud. This approach is aimed at reducing the sample size of the original data, while maintaining the available information on the curvature of the underlying manifold of interest. Specifically, we consider $D$-dimensional observations $\left\{x_{it}\right\}_{i=1, t=1}^{I_t, T}$ where the sample size $I_t$ is large. To reduce the sample size $I_t$, we use the high-dimensional mixture density estimation (HDMDE) method \citep[see][alg. 1]{mengPrincipalManifoldEstimation2021} to estimate a reduced set of points in the $D$-dimensional space that represent the shape of the data cloud $\left\{x_{it}\right\}_{i=1, t=1}^{I_t, T}$. Briefly, this approach is based on clustering of the data at each original time point to obtain cluster centers, $\left\{\mu_{j, t}\right\}_{j=1, t=1}^{N_t, T}$, represented using diamond-shaped points in Figure \ref{fig:lpme_step2}, along with their corresponding weights, $\{\hat{\theta}_{j, t}\}_{j=1, t=1}^{N_t, T}$, which characterize the shape of the original data. The number $N_t$ of clusters is estimated for each time point $t$ and is allowed to vary among time points to accommodate potential changes in the shape of the underlying structure. By implementing this approach, we reduce the size $\sum_{t=1}^T I_t$ of data $\left\{x_{it}\right\}_{i=1, t=1}^{I_t, T}$ to $\sum_{t=1}^T N_t$, where each $N_t$ tends to be much smaller than the corresponding $I_t$ \citep[see the simulation study presented in][fig. 3(c)]{mengPrincipalManifoldEstimation2021}. The centers $\left\{\mu_{j, t}\right\}_{j=1, t=1}^{N_t, T}$ estimated in this step are then used to reach an initial low-dimensional parameterization.

\subsubsection{Initialization}

The next step in the LPME algorithm for the estimation of the longitudinal principal manifold is initialization. The PME algorithm uses Isomap \citep{tenenbaumGlobalGeometricFramework2000} to develop an initial parameterization, which the PME algorithm proceeds to iteratively improve. Reaching a successful fit in the LPME algorithm requires consistent parameterizations across time points. Here, consistency refers to the similarity of the parameterizations of similarly shaped point clouds. However, using Isomap to find parameterizations of similarly shaped manifolds (i.e., the surfaces of brain regions observed at consecutive time points) is likely to result in inconsistent parameterizations. For example, the following two drastically different expressions parameterize the same curve: $x_1 = r$, $x_2 = \sin(r + \frac{\pi}{2})$ and $x_1 = -r$, $x_2 = \sin(\frac{\pi}{2} - r)$. Because Isomap tends to yield inconsistent parameterizations when applied to each time point separately, a different approach is needed for the LPME algorithm.

To obtain consistent parameterizations of the obtained cluster centers $\left\{\mu_{j, t}\right\}_{j=1}^{N_t}$ over time $t$, we propose using the PME algorithm to develop a parameterization of the centers $\left\{\mu_{j, t_1}\right\}_{j=1}^{N_{t_1}}$ taken from the first time point $t_1$, then using the projection index associated with this estimate to find initial parameterizations of the centers for the remaining time points. This approach is based on the assumption that the underlying manifolds of interest remains relatively stable over time.

\subsubsection{Fitting}

Once the initial parameterization for each cluster center $\mu_{j, t}$ is obtained, the model fitting process begins by using the cluster centers $\mu_{j, t}$, weights $\hat{\theta}_{j, t}$ and the parameterizations associated with each time point to initialize a PME run. Specifically, the cluster centers $\mu_{j, t}$ and weights $\hat{\theta}_{j, t}$ replace those obtained using the HDMDE algorithm, and the Isomap-derived parameters obtained in Steps 1 and 2 of Algorithm 2 in \cite{mengPrincipalManifoldEstimation2021}, respectively.
This results in an estimated principal manifold $\widehat{f_t}$ for each time point $t$ and an optimal tuning parameter $\lambda_t^*$ corresponding to the first two terms of the $\mathcal{K}_{\lambda, \gamma, \mathbb{P}}(F)$ defined in \eqref{eq:newKappa}, illustrated using lines in Figure \ref{fig:lpme_step3}. The performance of each fit $\widehat{f_t}$ is measured using $\tau_t = \frac{1}{I_t}\sum_{i=1}^{I_t}\| x_{it} - \widehat{f_t}(\pi_{\widehat{f_t}}(x_{it}))\|^{2}_{\mathbb{R}^D}$, an estimate of the mean squared distance between the observed data points and their projections to the fitted manifold $\widehat{f_t}$.

Further smoothing of the time-specific manifold estimates $\widehat{f_t}$ with respect to time is needed to consider curvature in the time dimension, corresponding to the third term of $\mathcal{K}_{\lambda, \gamma, \mathbb{P}}(F)$. As described in Section \ref{s:PME}, the principal manifold $\widehat{f_t}$ at time $t$ has spline form (equation~\eqref{eq: spline representation of the PME}) and is characterized by spline coefficients $\{s_{t, j, l}\}_{j=1, l=1}^{N, D}$ and $\{\alpha_{t, k, l}\}_{k=1, l=1}^{d+1, D}$. To smooth the functions $\widehat{f_t}$ over time $t$, we seek to estimate a spline function mapping from the time dimension to the coefficient space given a tuning parameter $\gamma$. As detailed by \cite{greenSilverman1994}, smoothing splines are fitted at a collection of knots corresponding to the design points (in this case, the cluster center parameterizations $\{\pi_{\widehat{f_t}}(\mu_{j, t})\}_{j=1, t=1}^{N_t, T}$). Because time-specific principal manifold estimates were fit with different cluster centers $\mu_{j, t}$ for each time point $t$, the coefficients of fitted functions at different time points are not comparable. Hence, to obtain coefficients that can be compared across time points, we design a grid of knots $\left\{\mathbf{r}_i\right\}_{i=1}^{N} \subset \mathbb{R}^d$ ranging from the minimum to the maximum of the approximated parameters $\{\pi_{\widehat{f_t}}(\mu_{j, t})\}_{j=1}^{N_t}$ in each dimension of $\mathbb{R}^d$. Define $N:=\max_{t=1,2,\ldots,T} N_t$ and $Y_{i,t}=\widehat{f_t}(\mathbf{r}_i)$ for all $i=1,2,\ldots,N$ and $t=1,2,\ldots,T$. Following the discussion in Chapter 7 of \cite{greenSilverman1994}, for each time point $t$, comparable spline coefficients $\mathbf{s}_t^*$ and $\mathbf{\alpha}_t^*$ can then be calculated by minimizing the following function.
\begin{equation}\label{eq:splineL}
    \mathcal{L}(\widehat{f_t}) = (\mathbf{Y}_t - \mathbf{E}\mathbf{s}_t - \mathbf{R}^\T\mathbf{\alpha}_t)^\T(\mathbf{Y}_t - \mathbf{E}\mathbf{s}_t - \mathbf{R}^\T\mathbf{\alpha}_t) + \lambda_t^*\mathbf{s}_t^\T \mathbf{E}\mathbf{s}_t,
\end{equation}
where $\mathbf{Y}_t$ is the $N \times D$ matrix consisting of elements $\left\{Y_{i,t}\right\}_{i=1}^{N}$ at each time point $t$; the matrix $\mathbf{E}=(E_{ij})_{1\le i,j\le N}$ is defined by $E_{ij} = \eta_{d}(\|\mathbf{r}_i - \mathbf{r}_j\|)$; $\eta_{d}(r) = r^{4 - d}\log r$ if $d$ is even and $\eta_d(r) = r^{4-d}$ if $d$ is odd; the functions $\phi_1,\ldots,\phi_{d+1}$ form a basis of the linear space of polynomials on $\mathbb{R}^d$ with degrees $\le1$; and the matrix $\mathbf{R}=(R_{ij})_{1\le i\le d+1,1\le j\le N}$ is defined by $R_{ij} = \phi_i(\mathbf{r}_j)$. By computing the derivative of $\mathcal{L}(\widehat{f_t})$ in (\ref{eq:splineL}) with respect to the coefficients, we conduct the minimization of $\mathcal{L}(\widehat{f_t})$ by solving the following system of equations.
\begin{equation}
  \left(
    \begin{array}{cc}
      \mathbf{E} + \lambda_t^* \mathbf{I} & \mathbf{R}^\T \\
      \mathbf{R} & \mathbf{0}
    \end{array}
  \right)\left(
    \begin{array}{c}
      \mathbf{s}_t \\
      \mathbf{\alpha}_t
    \end{array}
  \right) = \left(
    \begin{array}{c}
      \mathbf{Y}_t \\
      \mathbf{0}
    \end{array}
  \right). \label{eq:13}
\end{equation}
The solution to equation~\eqref{eq:13} , denoted by $\mathbf{\alpha}^*_t=(\alpha^*_{t,1},\ldots,\alpha^*_{t,d+1})^\T$ and $\mathbf{s}^*_t=(s^*_{t,1},\ldots,s^*_{t,N})^\T$, yields the function
$f_{t, l}^*(r) = \sum_{j=1}^{N}s_{t, j, l}^* \cdot  \eta_{d}\left(\|r - r_j^*\|\right) + \sum_{k=1}^{d + 1}\alpha_{t, k, l}^* \cdot \phi_k(r).$

We denote the manifold coefficients at each time point $t$ as $\mathbf{b}_t = (\mathbf{s}_t^{*\T}, \mathbf{\alpha}_t^{*\T})^\T$. To limit the ability of a poorly-fitting PME estimate to greatly influence the results of the LPME model, we fit a weighted spline model to smooth over the manifold coefficients $\mathbf{b}_t$, where the weights for each time point $t$ are equal to $w_t = 1 / \left(\tau_t\sum_{i=1}^T(1/\tau_i)\right)$. Thus, the weights correspond to the normalized inverse errors of the PME estimates at each time point. With a given tuning parameter $\gamma$, we then fit a weighted cubic spline function using $\left\{t_i\right\}_{i=1}^T$ as predictors and $\left\{\mathbf{b}_t\right\}_{t=1}^T$ as the response values, with weights $\left\{w_t\right\}_{t=1}^T$ resulting in a function taking the form
$g_{\gamma}(t) = \sum_{i=1}^{T}\delta_i \,\|t - t_i\|^{3} + \sum_{j=1}^{2}\nu_j\,\phi(t)_j$,
where by defining the $T \times T$ matrix $\mathbf{A}$ as  $\mathbf{A}_{ij} = \|t_i - t_j\|^{3}$ with $i,j = 1,\ldots, T$, the matrix $\mathbf{T}$ as  $\mathbf{T}_{ij} = \phi_i(t_j)$, and $\mathbf{W} = \operatorname{diag}(w_1, \dots, w_T)$, we obtain the coefficients $\mathbf{\delta}=(\delta_1,\ldots,\delta_T)^\T$ and $\mathbf{\nu}=(\nu_1,\nu_2)^\T$ as the solutions to
\begin{equation}
  \left(
  \begin{array}{ccc}
    2\mathbf{A}\mathbf{W}\mathbf{A} + 2\gamma\mathbf{A} & 2\mathbf{A}\mathbf{W}\mathbf{T} & \mathbf{T} \\
    2\mathbf{T}^\T\mathbf{W}\mathbf{A} & 2\mathbf{T}^\T\mathbf{W}\mathbf{T} & \mathbf{0} \\
    \mathbf{T}^\T & \mathbf{0} & \mathbf{0}
  \end{array}
  \right)\left(
  \begin{array}{c}
    \mathbf{\delta} \\
    \mathbf{\nu} \\
    \mathbf{m}
  \end{array}
  \right) = \left(
  \begin{array}{c}
    2\mathbf{A}\mathbf{W}\mathbf{B} \\
    2\mathbf{T}^\T\mathbf{W}\mathbf{B} \\
    \mathbf{0}
  \end{array}
  \right), \label{eq:16}
\end{equation}
where $\mathbf{B}=(\mathbf{b}_1,\ldots, \mathbf{b}_T)^\T$. For a given $\gamma$, we denote the manifold coefficients at time $t$ estimated by function $g(\cdot)$ as $\mathbf{B}_{\gamma}(t) = \left(\mathbf{s}_{\gamma}(t)^\T, \mathbf{\alpha}_{\gamma}(t)^\T\right)^\T$. Hence, given $\gamma$, the estimated embedding function at time $t$ and $d$-dimensional parameterization $\mathbf{r}$ is
\begin{equation}
  F_{\gamma}(t, \mathbf{r}) = \sum_{j=1}^{N}\mathbf{s}_{\gamma}(t)_j \eta_{d}\left(\|\mathbf{r} - \mathbf{r}_j^*\|\right) + \sum_{k=1}^{d+1}\mathbf{\alpha}_{\gamma}(t)_k \phi_k(\mathbf{r}). \label{eq:17}
\end{equation}

\subsubsection{Tuning}

The optimal tuning parameter $\gamma^*$ is identified using leave-one-out cross validation. In this process, the coefficient smoothing function is computed while excluding all data and coefficients with time $t$ from consideration, for $t = 1, \dots, T$. We denote this function by $g_{\gamma}^{(t)}$. We can then assess the performance of the LPME model using tuning value $\gamma$ by averaging over estimates of MSD calculated using the original observations associated with the left-out time $t$ as follows.
\begin{equation}
  \operatorname{MSD}(\gamma) = \frac{1}{T} \sum_{t=1}^{T}\frac{1}{I_t}\sum_{i=1}^{I_t}\|x_{i, t} - F_{\gamma}^{(t)}(t, \pi_{f_{\gamma}^{(t)}}(x_{i, t}))\|^2. \label{eq:18}
\end{equation}
The optimal tuning value $\gamma^*$ is chosen as the value of $\gamma$ that minimizes $MSD(\gamma)$. Leave-one-out cross validation was chosen for this application due to the relatively limited number of time points available for individuals in the ADNI dataset. In other circumstances with greater data availability, alternative cross validation methods, such as $k$-fold cross validation, could easily be used in place of leave-one-out cross validation to better meet computational demands.

In applying the proposed algorithms in extensive simulations, we did not encounter any issues with convergence of the algorithms. Further work must be done to demonstrate that the proposed estimator does indeed minimize the objective function, however \cite{mengPrincipalManifoldEstimation2021} demonstrated that the function that minimizes equation \eqref{eq:pme_kappa} is of smoothing spline form, while smoothing splines can be derived as a Bayes estimator in a penalized regression setting \citep{wahba1990}. These results motivate our expectation that the estimator from the LPME algorithm will minimize equation \eqref{eq:newKappa}.

\RestyleAlgo{ruled}
\LinesNumbered

\SetKwComment{Comment}{/* }{ */}

\RestyleAlgo{ruled}
\LinesNumbered

\begin{algorithm}
\caption{Longitudinal Principal Manifold Estimation (LPME)}\label{alg:lpme}
  \KwData{Data points $\left\{X_{i, t}\right\}_{i=1, t=1}^{I_t, T}$, positive integer $d$, positive integer $N_0 < \min{I_t} - 1$, $\alpha$, $\epsilon$, $\epsilon^* \in (0, 1)$, candidate tuning parameters $\left\{\lambda_k\right\}_{k=1}^K$, $\left\{\gamma_l\right\}_{l=1}^L$, $itr \geq 1$, which is the maximum number of iterations allowed.}
\KwResult{Analytic formula of $\hat{f}^*: \mathbb{R}^{d + 1} \to \mathbb{R}^{D + 1}$, optimal tuning parameter $\gamma^*$.}
\For{$t = 1, 2, \dots, T$} {
  Apply HDMDE algorithm with input $\left(\left\{X_{i, t}\right\}_{i = 1}^{I_t}, N_0, \epsilon, \alpha\right)$ and obtain $N_t$, $\left\{\mu_{j, t}\right\}_{j = 1}^{N_t}$, and $\left\{\theta_{j, t}\right\}_{j = 1}^{N_t}$\;
}

  Apply PME to parameterize $\left\{\mu_{j, 1}\right\}_{j = 1}^{N_1}$ by the $d$-dimensional parameters $\left\{r_{j, t}\right\}_{j = 1}^{N_1}$ and use $\pi_{f_1(0)}$ to parameterize $\left\{\mu_{j, t}\right\}_{j=1, t = 1}^{N_t, T}$ by $\left\{r_{j, t}\right\}_{j=1, t=2}^{N_t, T}$. Formally set $\pi_{f_1(0)}(\mu_{j, 1}) \gets r_{j, 1}$ for $j = 1, 2, \dots, N_1$, and $\pi_{f_{t}(0)}(\mu_{j, t}) \gets \pi_{f_1(0)}(\mu_{j, t})$ for $j = 1, 2, \dots, N_t$, $t = 2, \dots, T$\;

\For{t = 1, 2, \dots, T} {
  Apply modified PME algorithm with $\pi_{f_t(0)}(\mu_{j, N_t, t}) \gets \left\{r_{j, t}\right\}_{j = 1}^{N_t}$ and obtain $f_t$, $\lambda_t^*$, and $\tau_t$\;
  $\left\{r_{j, t}\right\} \gets \pi_{f_t}(\mu_{j, t})$ for $j = 1, \dots, N_t$ and $Y_{j, t} \gets f_t(r_{j, t})$ for $j = 1, \dots, N_t$\;
}

  Let $N = \max(N_t)$, $\left\{r_i^*\right\}_{i=1}^{N}$ be a grid spanning the range of estimated values of $\left\{r_{i, j}\right\}_{i=1, t=1}^{N_t, T}$\;

\For{t = 1, 2, \dots, T} {
  Set $Y_{i, t} = f_t(r_{i}^*)$ for $i = 1, \dots, N$\;
  Compute $f_t^*(r)$ by solving \eqref{eq:13}\;
  Set $w_t = \frac{1}{\tau_t \sum_{i=1}^{T}\frac{1}{\tau_i}}$\;
}

  Define $\mathbf{\omega}$ by setting $\mathbf{\omega}_t = \left[\mathbf{s}_t, \mathbf{\alpha}_t\right]$\;
  \For{l = 1, 2, \dots, L} {
    Compute $g_{\gamma_l}(t)$ by solving \eqref{eq:16}\;
    \For{t = 1, 2, \dots, T} {
      Compute $g_{\gamma_l}^{(t)}$ by solving \eqref{eq:16}\;
    }
    Estimate $MSD(\gamma_l)$ using \eqref{eq:18}\;
  }
  $\gamma^* = \arg\min_{\gamma}MSD(\gamma)$\;
  $f^*(t, \mathbf{r}) = f_{\gamma^*}(t, \mathbf{r})$, where the form of $f^*(t, \mathbf{r})$ is given in \eqref{eq:17}
\end{algorithm}

\subsection{Self-intersecting Manifold Estimation}\label{ss:selfInt}

Notably, many manifold learning methods are developed under the assumption that the underlying manifold is not self-intersecting. This is not the case in our motivating problem, where we are interested in modeling the surface of a brain region that is a closed 2-dimensional surface embedded in 3-dimensional space. Self-intersecting manifolds may appear in other applications where manifold learning is implemented. One such simple example is modeling digits, where the number 8, for instance, is a self-intersecting manifold.

The self-consistency condition used to define principal curves in \cite{hastiePrincipalCurves1989} theoretically precludes the estimation of curves that are self-intersecting. However, Hastie and Stuetzle demonstrated that this can be done empirically through the use of periodic smoothers. \cite{banfieldIceFloeIdentification1992} generalized the algorithm proposed in \cite{hastiePrincipalCurves1989}, averaging over projection residuals rather than data points, to fit closed curves. These approaches do not adapt easily to the higher dimensional penalized regression setting we consider here.

\begin{figure}
  \centering
  \subfloat[\centering Unaugmented Unit Circle]{{\includegraphics[width=7cm]{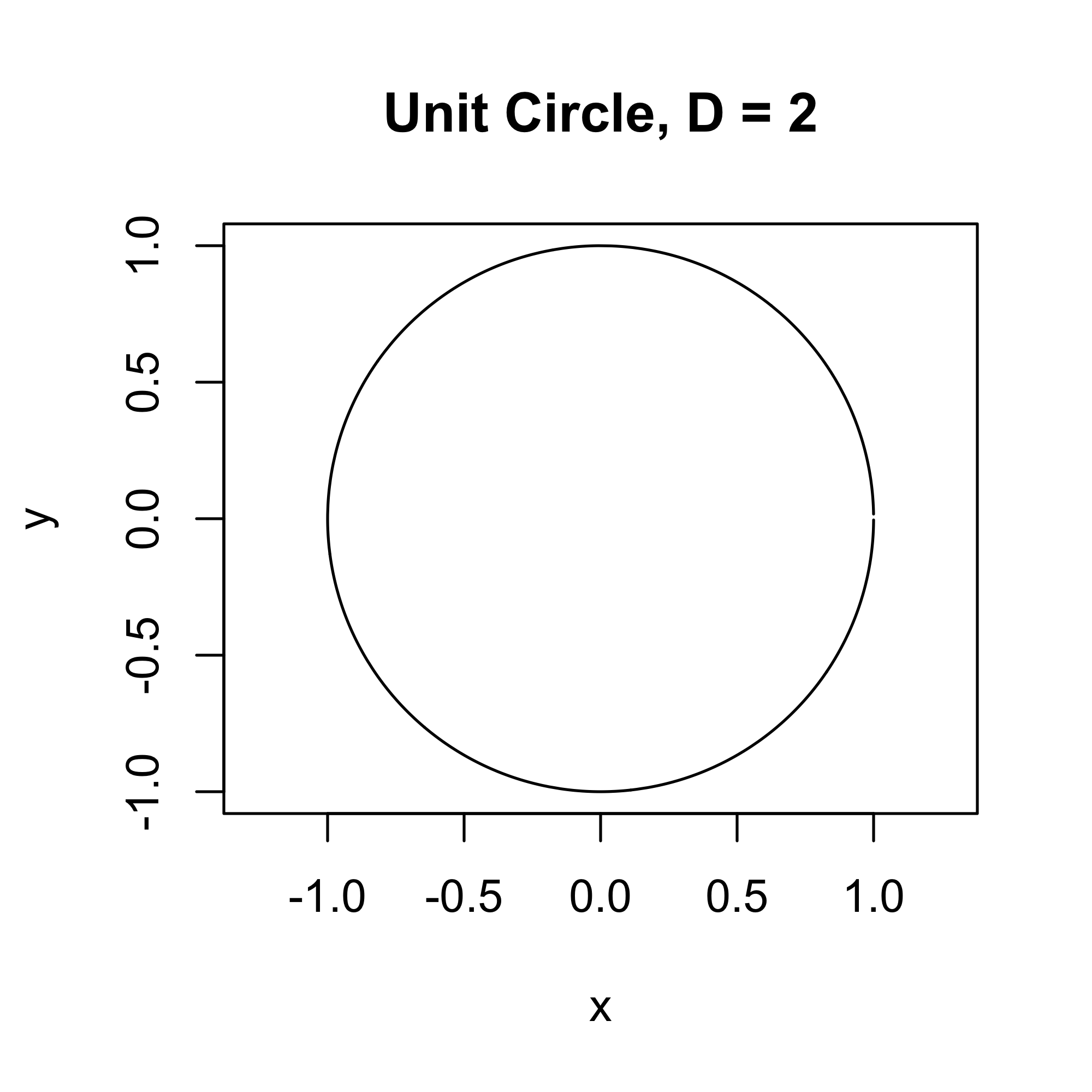}}}%
  \hfill
  \subfloat[\centering Augmented Unit Circle]{{\includegraphics[width=7cm]{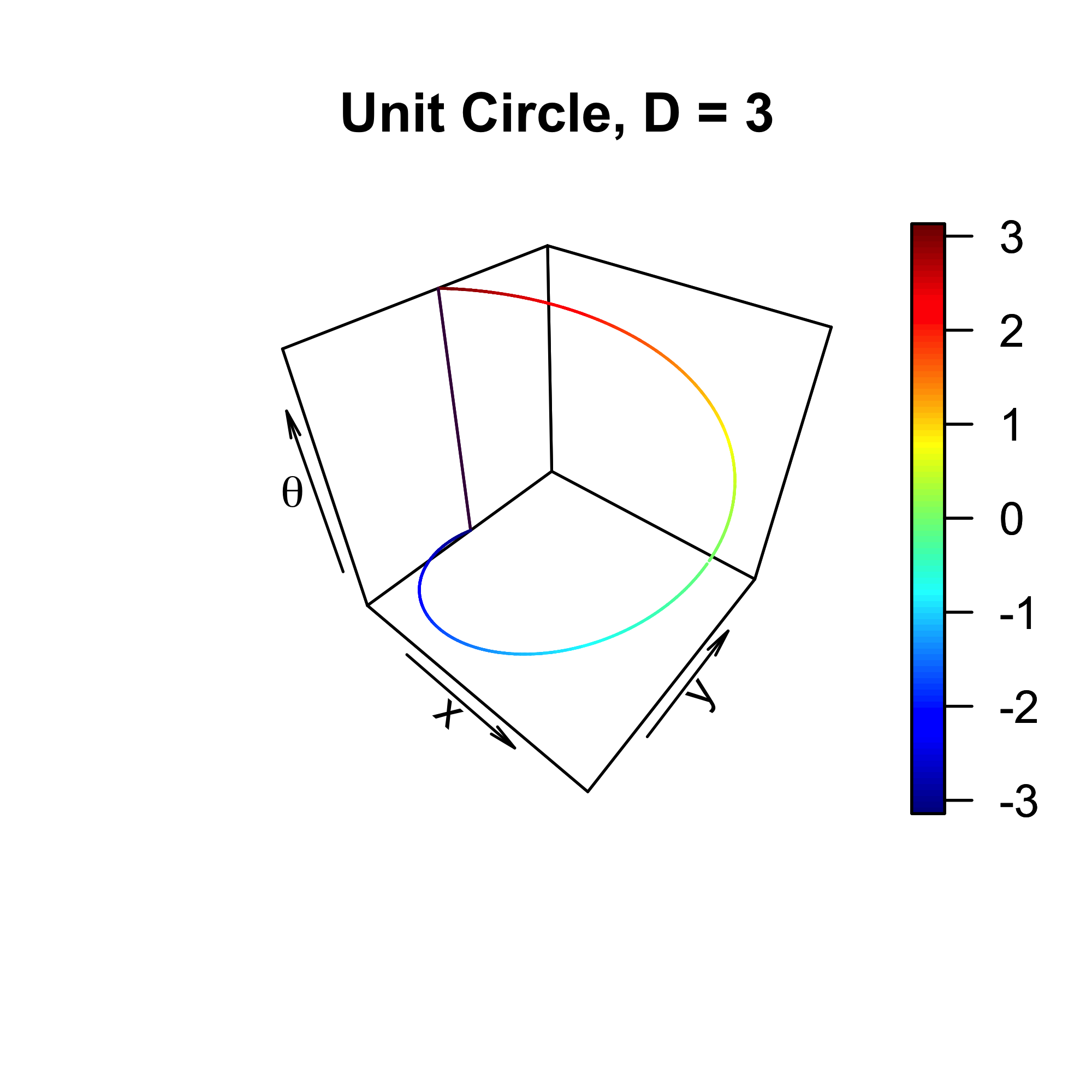}}}
  \caption{{\footnotesize Illustration of our proposed data augmentation approach using the Unit Circle. The left panel (a) shows the depiction of the circle with $d=1$ and $D=2$, while the right panel (b) shows how our proposed use of considering this same manifold in $D=3$ by using the polar coordinates results in a manifold that is not self-intersecting.}}
  \label{fig:unit_circle_augmentation}
\end{figure}

A commonly used alternative approach for handling closed manifolds is partitioning the high-dimensional data cloud into parts that are not self-intersecting, performing manifold fitting for each partition, and then combining the results, preferably by using a technique that results in a smooth manifold in the low-dimensional space \citep[e.g.][]{mengPrincipalManifoldEstimation2021}. To avoid this process, we propose a data augmentation approach taking advantage of polar or spherical coordinates, depending on the value of $D$. This approach is motivated by the concept of ``lift" in algebraic topology \citep[sec. 1.1]{hatcher2002algebraic}. As a simple example, the unit circle can be considered a self-intersecting manifold with $d = 1$ when viewed in two dimensions. However, by adding a third dimension equal to the angle of each observation from the origin, the manifold can be viewed as given in the space with $d = 1$ with the data cloud of $D = 3$, while avoiding any self-intersections. This enables the PME algorithm, and thus the LPME algorithm, to be fit under these circumstances. This process is illustrated graphically in Figure \ref{fig:unit_circle_augmentation}.

Because the added dimensions duplicate information contained within the unaugmented observations, these dimensions can be discarded to reach estimates in the original $D$-dimensional space. This approach to data augmentation offers an alternative to manual partitioning used by \cite{mengPrincipalManifoldEstimation2021}. Reaching a combined estimate after fitting PME to each partition individually requires ``gluing'' the estimated manifolds on each partition together, which may introduce error to the combined manifold estimate. This source of error is avoided using our proposed data augmentation approach. Additionally, scaling the added dimensions by a constant value enables the modification of the level of curvature present in the manifold. This may encourage improved performance when estimating manifolds that have too much curvature for PME to reach an appropriate fit under normal circumstances. However, an in depth comparison of these approaches is beyond the scope of this article and is left for future work.

\section{Simulations}\label{s:simulations}

To assess the performance of the LPME algorithm (Algorithm \ref{alg:lpme}), we use simulation studies to compare how closely (in terms of reduction of MSE) the LPME-estimated underlying functions approximate the true data generating functions in comparison with that of three alternative approaches. We consider settings in which $d = 1$ and $D = 2$, $d = 1$ and $D = 3$, as well as $d = 2$ and $D = 3$, with several manifolds being used to generate datasets in each setting. For each manifold, we consider differing values for the time duration, interval between observations, noise levels within and between time points, and the type and magnitude of structural changes in the underlying manifold over time. The embedding functions used to define each manifold are given in Table \ref{table:simulation_embeddings}. Based on our observations of systematic noise sources in imaging data, we add various types of terms to these functions at each visit. In these functions, the value of $\zeta$ and the functional form of $g(\cdot)$ are varied between visits and represent structural changes in the underlying manifold over time. The function $g(\cdot)$ may show change with respect to time that is constant (no change), linear, quadratic, or sinusoidal, with $\zeta$ serving as a scalar multiplier. The values of $\iota$ represent within-image noise in the high-dimensional space, and are randomly sampled from a normal distribution with mean zero and variance that is varied between visits. Meanwhile, $\alpha$ and $\beta$ are vectors of length $d$ that are drawn from a normal distribution with mean 1 and variance that is varied between visits. These values describe random fluctuations in the manifold between time points, reflecting noise introduced by different imaging sessions.

\begin{table}[ht]
  \centering
  \resizebox{\textwidth}{!}{\begin{tabular}{|c c c c c|}
    \hline
    Case & $d$ & $D$ & $f_t(\mathbf{r})$ & Domain \\
    \hline
    1 & 1 & 2 & $\left(r, \ \alpha \text{sin} \ (\beta r + \frac{\pi}{2})\right) + \zeta\mathbf{g}(t) + \iota$ & $-3 \leq r \leq 3$ \\
    2 & 1 & 2 & $\left(r, \ \alpha\text{sin} \ (\beta r)\right) + \zeta\mathbf{g}(t) + \iota$ & $-3\pi \leq r \leq 3\pi$ \\
    3 & 1 & 2 & $\left(\alpha_1 \text{cos} \ (\beta_1 r), \ \alpha_2\text{sin} \ (\beta_2 r)\right) + \zeta\mathbf{g}(t) + \iota$ & $-\frac{4\pi}{5} \leq r \leq \frac{\pi}{2}$ \\
    4 & 1 & 3 & $\left(r, (\alpha_1r + \beta_1)^2, \ (\alpha_2r + \beta_2)^3\right) + \zeta\mathbf{g}(t) + \iota$ & $-1 \leq r \leq 1$ \\
    5 & 1 & 3 & $\left(r, \ \alpha_1\text{cos} \ (\beta_1 r), \ \alpha_2\text{sin} \ (\beta_2 r) \right) + \zeta\mathbf{g}(t) + \iota$ & $0 \leq r \leq 3\pi$ \\
    6 & 2 & 3 & $\left(\beta_1r_1, \ \beta_2r_2, \ \alpha_1(\alpha_2\|\beta\mathbf{r}\|^2)\right) + \zeta\mathbf{g}(t) + \iota$ & $-1 \leq r_1, r_2 \leq 1$\\
    7 & 2 & 3 & $\left(\alpha_1\beta_1r_1\text{cos} \ (\alpha_1r_1), \ \alpha_2\beta_2r_1\text{sin} \ (\alpha_2r_1), \ r_2\right) + \zeta\mathbf{g}(t) + \iota$ & $0 \leq r_1 \leq 3\pi$; \ $-1 \leq r_2 \leq 1$\\
    8 & 2 & 3 & $\left(\alpha_1\text{sin}(\beta_1r_1)\text{cos}(\beta_2r_2), \ \alpha_1\text{sin}(\beta_1r_1)\text{sin}(\beta_2r_2), \ \alpha_1\text{cos}(\beta_1r_1)\right) + \zeta\mathbf{g}(t) + \iota$ & $0 \leq r_1 \leq \pi$; \ $0 \leq r_2 \leq 2\pi$\\
    \hline
  \end{tabular}}
  \caption{{\footnotesize Embedding functions used for simulation studies. Function $\bf{g}(t)$ denotes structural change in the underlying manifold over time, while $\zeta$ represents the scale of this change. Parameters $\alpha$ and $\beta$ are normally distributed with mean 1, and represent random changes in the manifold between time points, while $\iota$ represents within-image noise.}}
  \label{table:simulation_embeddings}
\end{table}

In the situations where $d = 1$, LPME is compared to the PME method naïvely run at each time point without smoothing over time, as well as the principal curve algorithm described in \cite{hastiePrincipalCurves1989}, also run independently at each time point. The principal curve algorithm is implemented using the \texttt{principal\_curve()} function in the \texttt{princurve} package \citep{Cannoodt2018princurve}, developed using \texttt{R} \citep{rSoftware2023}. In this function, each of the three smoothing options, \texttt{smooth\_spline}, \texttt{lowess}, and \texttt{periodic\_lowess}, are tested, with the option resulting in the lowest error from the true values being chosen. Following \cite{mengPrincipalManifoldEstimation2021}, the inputs for the PME algorithm are set to $\alpha = 0.05$, and $\epsilon = 0.001$, with $\lambda_g = \exp(g)$ for $g = -15, \dots, 5$. The minimum number of cluster centers is set to $10 \times d$. In cases where $d = 2$, LPME is again compared to the naïve PME approach described above, and the principal surface estimation algorithm developed by \cite{yueParameterizationWhiteMatter2016}.

A factorial design is used to run the simulation studies, with the factor levels set as follows: 1) $\alpha, \beta, \zeta \in \left\{0, 0.05, 0.1, 0.25, 0.5, 1.0\right\}$; 2) study duration: $\left\{1, 2, 5\right\}$; 3) interval between images: $\left\{0.1, 0.25, 0.5\right\}$; 4) longitudinal change model: Constant, Linear, Quadratic, Sinusoidal. The sample size for each time point is set to 1,000 observations. Each combination of factor levels is run once, resulting in each embedding map being simulated a total of 7,776 times. \if1\blind{All code required to reproduce the simulation results and associated Figures is provided at \href{https://github.com/rjzielinski/lpme-project}{\texttt{https://github.com/rjzielinski/lpme-project}}.}\fi

\begin{figure}
  \centering
  \includegraphics[height=9cm]{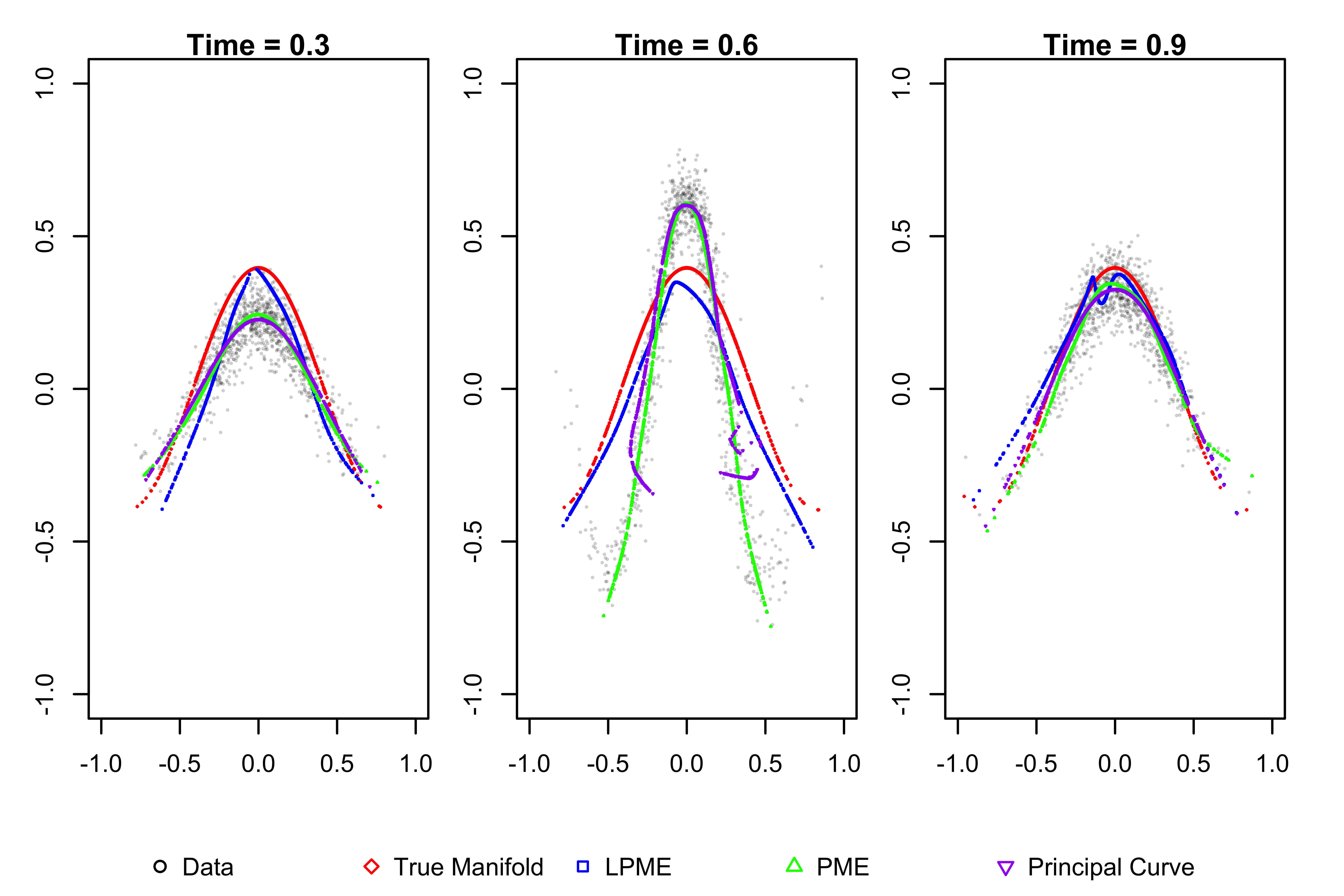}
  \caption{{\footnotesize Simulation Case 1. Scatterplots of one example simulated data cloud (circular points) with $d=1$ overlaid with points on the true embedding manifold (diamonds), the estimated LMPE function (squares), PME (triangles), and principal curve or surface (inverted triangles).}}
  \label{fig:sim_case1}
\end{figure}

Visualizations of the results from one example simulated case, truncated for concision, are shown in Figure \ref{fig:sim_case1}. Informally, the performance of each estimation method can be observed by considering the proximity of each method's estimated manifold to the true manifold over time. Thus, in Figure \ref{fig:sim_case1}, we see that the LPME-estimated manifold bears the closest resemblance to the true manifold underlying the data, while the manifolds estimated by PME and the principal curve method show greater responses to temporary random fluctuations in the data at each time point.

\begin{table}[h]
  \centering
  \begin{tabular}{|c c c c c|}
    \hline
    Case & Data & LPME & PME & PC/PS \\
    \hline
    1 & 0.146 (0.233) & {\bf 0.074 (0.122)} & 0.131 (0.258) & 0.118 (0.206) \\
    2 & 0.467 (0.665) & {\bf 0.248 (0.516)} & 0.516 (0.750) & 0.564 (0.419) \\
    3 & 0.291 (0.601) & {\bf 0.239 (0.564)} & 0.317 (0.640) & 0.264 (0.584) \\
    4 & 4.26 (21.2) & {\bf 3.29 (13.0)} & 4.23 (21.1) & 4.22 (21.2) \\
    5 & 0.895 (1.33) & {\bf 0.584 (1.08)} & 0.894 (1.36) & 0.821 (1.22) \\
    6 & 0.284 (1.06) & {\bf 0.273 (0.891)} & 0.316 (1.04) & 0.557 (0.387) \\
    7 & 0.145 (0.552) & 2.96 (4.65) & 6.92 (1.17) & {\bf 1.58 (0.514)} \\
    8 & 0.110 (0.325) & {\bf 0.074 (0.208)} & 0.115 (0.331) & 0.172 (0.234) \\
    \hline
  \end{tabular}
  \caption{{\footnotesize Mean Squared Distance comparison of simulated data, longitudinal principal manifold estimation (LPME)-, principal manifold estimation (PME)-, and principal curve / surface-based estimates to true values, Median (IQR). The lowest algorithm-specific median (IQR) are highlighted in bold.}}
  \label{table:simulation_results_median}
\end{table}

The median and interquartile range of the mean squared distance from the true underlying manifold values for the estimates of each approach, as well as for the data itself, are shown in Table \ref{table:simulation_results_median}. Analogous mean and standard deviation summaries of the results are given in Table S1 of the online supplement. Because the PME, principal curve, and principal surface methods each attempt to estimate the manifold in question at each individual time point without allowing other time points to inform these estimates, they should result in similar deviations from the true underlying manifold, with differences in error resulting primarily from differing performances in fitting to the observed data. Meanwhile, because the LPME approach accounts for all time points simultaneously, this approach results in lower mean squared distance values.

The result summaries indicate that in most cases, LPME provides a substantial improvement in performance over the PME and principal curve approaches when estimating the underlying manifold. As seen clearly in Figure \ref{fig:sim_case1}, while the structure of the simulated data changes noticeably between time points, the LPME estimates remain relatively stable over time. This contrasts to the estimates found by the PME and principal curve approaches, which, as expected, are highly sensitive to the added systematic noise in the data observed at each given time point. This ultimately results in the LPME estimates remaining closer to the true underlying manifold. Similar results hold in most of the other simulation cases.

\section{Longitudinal Segmentation of MRI Images}\label{s:application}

As discussed previously, meaningful between-image errors in estimates of subcortical structures are introduced during the process of segmenting MRI images. This section demonstrates how the LPME algorithm may be used to mitigate the effects of such noise on further analysis using these estimates. To achieve this, we used MRI data collected through the ADNI study, a longitudinal observational study with the goal of identifying imaging biomarkers to assess the progression of AD.

While the original ADNI 1 cohort included 200 cognitively healthy elderly individuals, 400 with mild cognitive impairment, and 200 with AD, this analysis focuses on 463 participants, of whom 130 were cognitively normal, 223 displayed symptoms of mild cognitive impairment, and 101 were diagnosed with AD at the baseline study visit. Diagnostic information was unavailable for 9 of the 463 participants for whom imaging data was available. Of the 130 cognitively normal participants, 34 were subsequently diagnosed with mild cognitive impairment or dementia during the course of the study. Follow-up duration for these participants ranged from 0 months to approximately 52 months, with imaging scheduled to be conducted at six- or 12-month intervals depending on the stage of the study. To encourage greater stability of longitudinal estimates, only participants with at least 24 months of follow-up were considered. Our final analysis set included 236 participants. Of these remaining participants, 88 were cognitively normal, 107 had mild cognitive impairment, and 41 were diagnosed with AD at their baseline visits. We focused on two brain regions of interest: the hippocampus and the thalamus. The hippocampus is of interest due to the changes experienced by those with AD, while the shape of the thalamus more closely aligns with the spherical structure used in simulation case 8.

Images were processed using FSL via the \texttt{fslr} package \citep{muschelliFslrConnectingFSL2015} in \texttt{R}, with FSL's FLIRT method linearly registering the images to MNI space, and the FIRST method being used for image segmentation. Following image segmentation, the surfaces of each region were identified by finding the extreme voxels in each dimension with nonzero intensity readings. The estimated surface positions were then standardized to a maximum distance of one from the origin in each dimension of Euclidean space and centered around the origin. Finally, the data were augmented with spherical coordinates in a manner similar to that described in Section \ref{ss:selfInt} to avoid the need to fit to a self-intersecting manifold.

Results of fitting the PME and LPME algorithms on the surface of the left hippocampus and left thalamus of a single participant are shown in Figure S1 in the online supplement. The previously described data augmentation approach was used to enable the fitting of closed surfaces. Visual inspection of the hippocampus estimates indicates that while there are slight differences between the shape shown in the data and the shape estimated by the LPME algorithm, the surface estimated by LPME appears to fit reasonably to the data. There are two main sources of discrepancies between the data and the LPME estimate. First, at time points where the observed surface changes orientation, the LPME estimates maintain a consistent orientation, reflecting the goal of encouraging stability in the structural estimates between time points. The second difference between the LPME estimates and the observed data is seen at the sharper corners of the hippocampus, where the LPME estimates do not fully capture the severity of curves in the observed data.

\begin{figure}
  \centering
  \subfloat[\centering Left Hippocampus Cross Section]{{\includegraphics[width=0.49\textwidth]{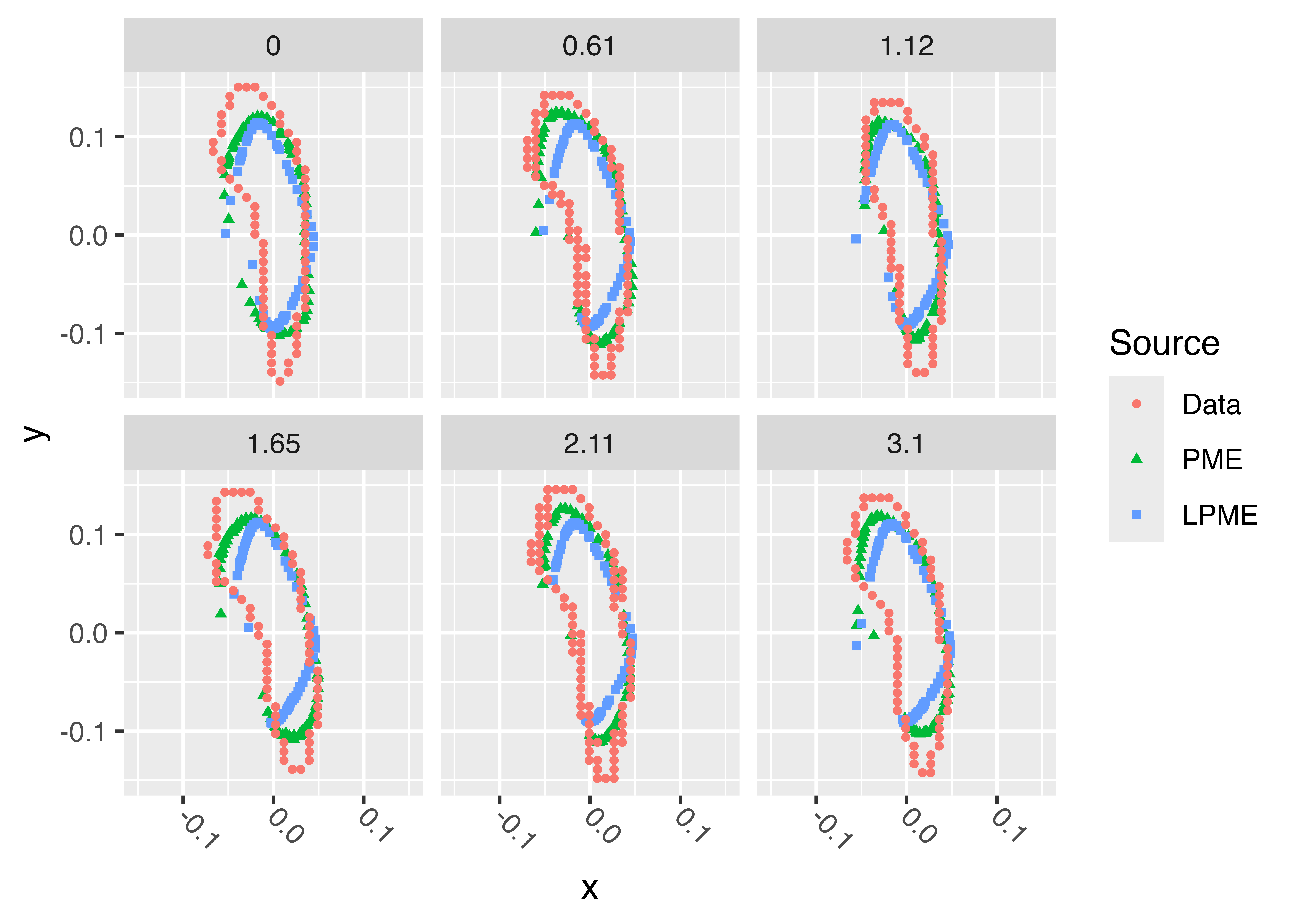}} \label{fig:lhipp_cross_sections}}%
  \hfill
  \subfloat[\centering Left Thalamus Cross Section]{{\includegraphics[width=0.49\textwidth]{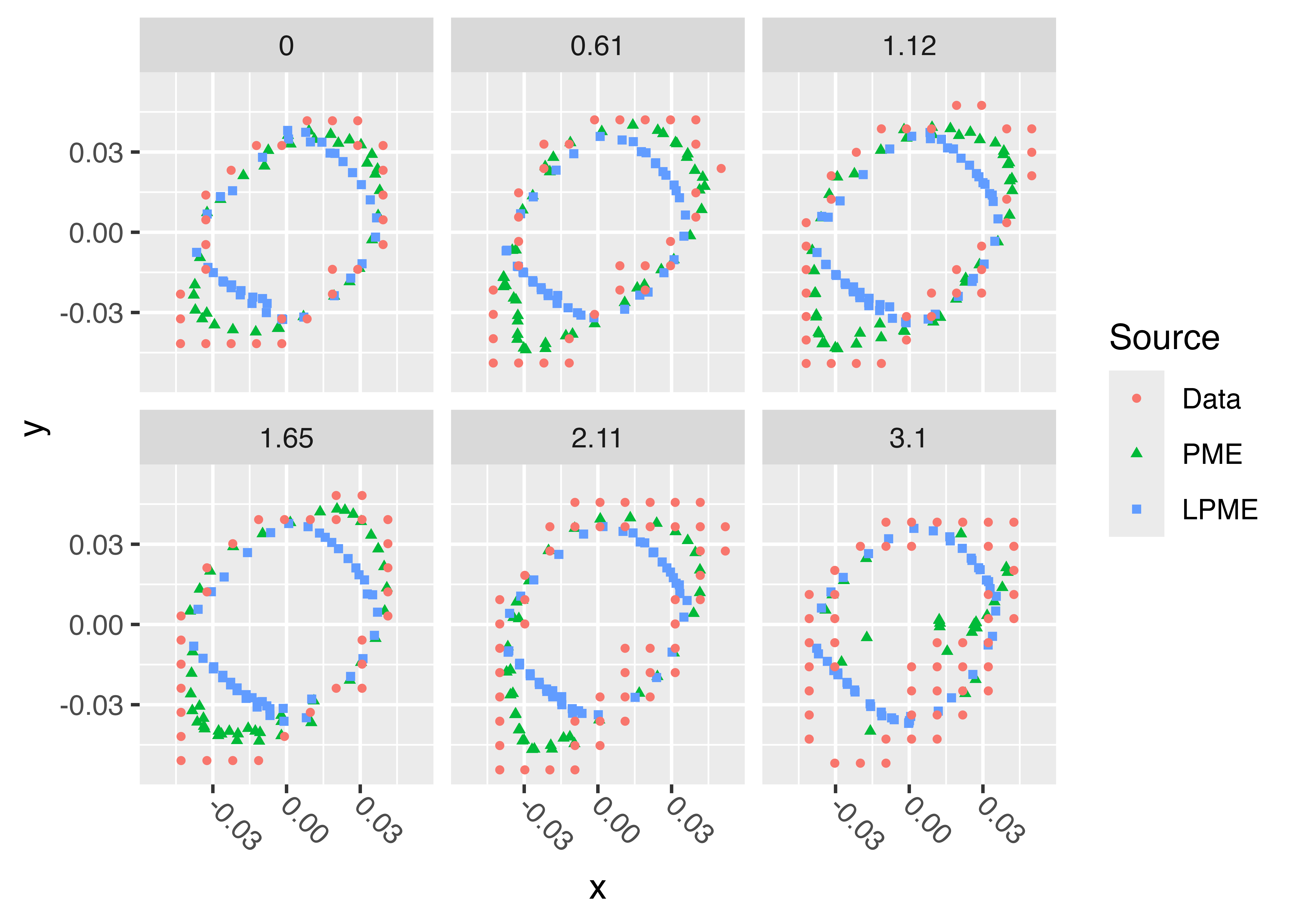}} \label{fig:lthal_cross_sections}}

  \caption{{\footnotesize Left Hippocampus and Left Thalamus Cross Sections. LPME-approximated embedding function routinely falls inside the boundaries of the observed hippocampus surface. In the fifth hippocampus observation with $t=2.11$, the LPME-estimated surface has a gap. Such gaps in the surface estimates frequently correspond with inaccurate volume estimates. The cross sectional plots indicate that the PME and LPME algorithms fit much more closely to the regularly shaped thalamus surface.}}
  \label{fig:adni_cross_sections}
\end{figure}

Figure \ref{fig:lhipp_cross_sections} depicts cross sections of the observed hippocampus surface and the estimated surface obtained using LPME and PME at each individual time point. These cross sections illustrate the differences between the estimates reached by the PME and LPME algorithms, with the LPME-based estimate being generally less responsive to changes in the shape and orientation of the hippocampus between time points. This Figure reiterates that both algorithms struggle to capture the distinctive sharp curves in the structure near the top and bottom corners of each plot. Additionally, we observe a gap in the estimated manifold along the lower-left edge of the surface. This is present for both the PME and LPME estimates, but appears to be more prevalent when using LPME. It also appears that the boundaries of the LPME-estimated surface tend to fall inside the boundaries of the PME-estimated surface, particularly at locations with high levels of curvature.

The hippocampus is a region with a relatively irregular shape, which may account for the differences in performance between the LPME algorithm when applied to simulated data and when applied to the hippocampus data. To understand how the approach performs when used with a more regularly shaped brain region, we also applied the PME and LPME algorithms to the surface of the left thalamus of the same set of individuals described previously. Estimates of the surface of the thalamus were obtained from the MRI images using the same preprocessing and segmentation steps detailed above for the hippocampus data. Considering the same individual shown in Figure \ref{fig:lhipp_cross_sections}, cross sectional results of fitting the PME and LPME algorithms to the surface of the left thalamus are shown in Figure \ref{fig:lthal_cross_sections}. The cross section plots shown in Figure \ref{fig:lthal_cross_sections} indicate that both the PME and LPME algorithms show a closer fit to the more regularly shaped thalamus surface than was achieved for the hippocampus, though the LPME estimate still falls inside the observed surface.

To understand how the subcortical structure estimates obtained by the LPME and PME algorithms impact the corresponding volume estimates for these structures, the volume of each structure was approximated from the raw data and the estimates obtained by the PME and LPME algorithms at each time point. The volume estimation approach used here relies on counting the voxels contained within the boundary defined by the PME- or LPME-estimated embedding map, with each voxel having a known volume. \if1\blind{The code used to implement this approach is available at \href{https://github.com/rjzielinski/lpme-project}{\texttt{https://github.com/rjzielinski/lpme-project}}.}\fi To align processing steps as closely as possible, the volume estimates shown for the raw data use the same volume estimation method used to find LPME- and PME-based approximations, rather than the volume estimates produced by FSL's FIRST method during segmentation.

The volume estimates for randomly selected individuals can be found in panels (a) and (b) of Figure S2 in the online supplement, corresponding to estimates for the left hippocampus and left thalamus, respectively. Here, we see that volume estimates differ substantially depending on whether PME, LPME, or the raw data were used, with the LPME estimate resulting in the lowest volume computations. In panel (a) of Figure S2, we see that the LPME-based volume estimates for the hippocampus are consistently lower than those obtained from the raw data and the PME estimates. All three sources show substantial variability between study visits, particularly for subjects \texttt{033\_S\_0514} and \texttt{099\_S\_0291}. Considering the thalamus estimates in panel (b), we again see that meaningful variability in the volume estimates using the raw data and PME estimates as sources exist between time points. However, the LPME estimates demonstrate the ability to smooth over this variability, resulting in volume estimates that are not as sensitive to differences between time points. We also see that when basing volume estimates on the more reliable LPME fit to the thalamus data, LPME yields volume estimates similar in magnitude to those obtained from the raw data and PME estimates.

\begin{table}
  \centering
  \begin{tabular}{|c c c c c|}
    \hline
    Structure & Regression-Adjusted & Data & LPME & PME  \\
    \hline
    Left Hippocampus & No & 129 & \textbf{119} & 124 \\
    Right Hippocampus & No & 133 & \textbf{124} & 179 \\
    Left Hippocampus & Yes & 104 & \textbf{93.2} & 98.4 \\
    Right Hippocampus & Yes & 107 & \textbf{100} & 149 \\
    Left Thalamus & No & 55.7 & \textbf{28.5} & 42.5 \\
    Right Thalamus & No & 108 & \textbf{61.7} & 93.3 \\
    Left Thalamus & Yes & 45.2 & \textbf{16.0} & 35.3 \\
    Right Thalamus & Yes & 86.9 & \textbf{42.7} & 76.6 \\
    \hline
  \end{tabular}
  \caption{{\footnotesize Median of the empirical standard deviation of volume estimates computed for each ADNI participant. Entries that were not regression-adjusted include computations for raw volume estimates, while regression-adjusted entries compute the standard deviations of linear regression models regressing the volume estimates on time.} }
  \label{table:adni_volume_sds}
\end{table}

A summary of the between-visit variability in the volume estimates of the subcortical structures considered here is shown in Table \ref{table:adni_volume_sds}. In addition to reporting the standard deviations of the volume estimates, we also report regression-adjusted estimates of variability to account for cases in which the structure cannot be assumed to be constant over time, as is the case for those with AD. The regression-adjusted summaries estimate the standard deviation of the residuals of a simple linear regression of volume on time. Meanwhile, in cases without regression adjustment, we estimated the standard deviation of the volume estimates for each individual and each estimate source (i.e. raw data, LPME, and PME). We then took the median of these values for each estimate source. Table \ref{table:adni_volume_sds} shows that, despite the visual evidence of inconsistency in the LPME estimates of the hippocampus caused by its irregular shape, LPME still yields the smoothest volume estimates for both hippocampuses. LPME also results in the smoothest volume estimates when applied to the thalamus, showing even greater improvements over the alternative volume estimates.

\section{Discussion}\label{s:discussion}

In this article, we propose the LPME framework for modeling longitudinal changes in a low-dimensional manifold underlying high-dimensional data. This work to adapt a nonlinear manifold learning method to a general-purpose longitudinal setting expands the situations in which such methods may be applicable. We also suggest a data augmentation approach that circumvents challenges in fitting principal manifold-based methods to self-intersecting manifolds in select settings. In simulated datasets where spurious changes in the structure to be estimated are introduced between time points, the LPME algorithm shows improved performance in recovering the underlying manifold when compared to a naïve approach of applying either the PME or principal curve/surface algorithms at each time point.

While the LPME algorithm demonstrated strong performance in a number of settings, it is important to note the assumptions under which the method performs well. First, because Isomap and PME are used for initialization and time-specific fitting, the assumptions required by these algorithms are also needed for the LPME algorithm to perform well. In practice this limits the success of LPME to situations where Isomap performs well, which excludes nonconvex manifolds, manifolds that have holes, or manifolds that have too high a level of curvature. The use of alternative methods, such as Locally Linear Embedding \citep{roweisNonlinearDimensionalityReduction2000} or Laplacian Eigenmaps \citep{belkin2003laplacian}, to generate the initial parameterization may alleviate this concern. A second assumption is that the LPME method requires relative stability in the overall shape of the underlying manifold structure. Thus, while this algorithm is useful for separating meaningful changes in a manifold over time from inter-observation error, it may not perform well when the error between observations is too substantial, similar to any smoothing method. How restrictive this requirement is depends largely on the area of application. When used to estimate subcortical surfaces in a single individual as suggested here, this tends not to be overly restrictive, as changes in the shape and size of a subcortical structure, while clinically meaningful, tend to be relatively small in magnitude compared to their defining features.

An interesting future research direction is extension of the LPME method for modeling  shapes such as the hippocampus surfaces obtained from the ADNI dataset. It is clear that neither the PME nor the LPME estimates were able to fully capture the distinctive curved structure of the hippocampus. There are two potential sources of this difficulty, each suggesting different steps that may be taken to address this issue. Replacing the simple grid search that is used to select the optimal value of the smoothing parameters in the PME and LPME algorithms with an improved search method could yield results that are more sensitive to the nuances of the structure being estimated, such as the areas of sharp curvature in the hippocampus.

Second, further investigation of the  data augmentation approach described in Section \ref{ss:selfInt} may enable fitting the PME or LPME algorithms to a more extensive set of structures. The proposed augmentation with polar or spherical coordinates allowed the PME and LPME algorithms to estimate self-intersecting manifolds that would not previously have been estimable by these methods. However, while this approach addressed a problem posed by structures with roughly circular or spherical shapes, such as the thalamus, its success may not extend to settings with more complex data. In the case of the hippocampus, potential intersections in the spherical coordinates may result in the overlaps in certain regions of the estimated manifolds. Similar types of issues may arise in settings with other complex structures, like a figure 8 shape. The development of a procedure to handle these considerations would expand the potential use cases for this approach, thus allowing principal manifold-based methods to be used in a wider range of situations.

Estimation of volumes of the resulting manifolds is essential, as the volumes are often used as outcomes in modeling disease processes over time in clinical trials and observational studies. In this manuscript, the volumes were obtained by simply computing the number of within-region voxels resulting from each fitting procedure. If the resulting manifold has gaps, however, this procedure may not be possible to implement. \cite{mengPrincipalManifoldEstimation2021} introduced an interior identification approach that effectively makes use of the explicit representation of the embedding map obtained through the PME and LPME fitting processes. Future work may adapt this procedure to settings with gaps in the estimated manifold.

While the approach described in this article is limited to the estimation of manifolds from observations taken from a single structure over time, extension of this method to manifold estimation for groups of similar structures will be a valuable contribution to the literature allowing population-level comparisons of time-dependent trajectories of shapes. Reaching estimates associated with groups of structures raises the possibility of statistical comparisons between these groups, which could prove beneficial in a number of clinical trial settings.

Finally, Gaussian process regression (GPR) may be applied to smooth the time direction. The smoothness can be represented by the covariance kernel used in the GPR, e.g. by applying the Gaussian or Matérn kernels \citep{li2023inference}. Future work may consider implementation of a similar approach in the framework we proposed in this article.

\if1\blind
{
\section*{Acknowledgments}

Research presented in this manuscript was funded by National Institute On Aging of the National Institutes of Health (NIH) under Award Number R01AG075511. The ideas presented in this paper are the responsibility of the authors and may not necessarily represent the official views of the NIH.

Data collection and sharing for the Alzheimer's Disease Neuroimaging Initiative (ADNI) is funded by the National Institute on Aging (National Institutes of Health Grant U19 AG024904). The grantee organization is the Northern California Institute for Research and Education. In the past, ADNI has also received funding from the National Institute of Biomedical Imaging and Bioengineering, the Canadian Institutes of Health Research, and private sector contributions through the Foundation for the National Institutes of Health (FNIH) including generous contributions from the following: AbbVie, Alzheimer’s Association; Alzheimer’s Drug Discovery Foundation; Araclon Biotech; BioClinica, Inc.; Biogen; Bristol-Myers Squibb Company; CereSpir, Inc.; Cogstate; Eisai Inc.; Elan Pharmaceuticals, Inc.; Eli Lilly and Company; EuroImmun; F. Hoffmann-La Roche Ltd and its affiliated company Genentech, Inc.; Fujirebio; GE Healthcare; IXICO Ltd.; Janssen Alzheimer Immunotherapy Research \& Development, LLC.; Johnson \& Johnson Pharmaceutical Research \& Development LLC.; Lumosity; Lundbeck; Merck \& Co., Inc.; Meso Scale Diagnostics, LLC.; NeuroRx Research; Neurotrack Technologies; Novartis Pharmaceuticals Corporation; Pfizer Inc.; Piramal Imaging; Servier; Takeda Pharmaceutical Company; and Transition Therapeutics.
} \fi

\newpage
\bibliographystyle{abbrvnat}
\bibliography{references}

\begin{thebibliography}{26}
\providecommand{\natexlab}[1]{#1}
\providecommand{\url}[1]{\texttt{#1}}
\expandafter\ifx\csname urlstyle\endcsname\relax
  \providecommand{\doi}[1]{doi: #1}\else
  \providecommand{\doi}{doi: \begingroup \urlstyle{rm}\Url}\fi

\bibitem[Banfield and Raftery(1992)]{banfieldIceFloeIdentification1992}
J.~D. Banfield and A.~E. Raftery.
\newblock Ice {{Floe Identification}} in {{Satellite Images Using Mathematical Morphology}} and {{Clustering}} about {{Principal Curves}}.
\newblock \emph{J. Amer. Stat. Assoc.}, 87\penalty0 (417):\penalty0 7--16, Mar. 1992.
\newblock ISSN 0162-1459.

\bibitem[Belkin and Niyogi(2003)]{belkin2003laplacian}
M.~Belkin and P.~Niyogi.
\newblock Laplacian {{Eigenmaps}} for {{Dimensionality Reduction}} and {{Data Representation}}.
\newblock \emph{Neural Computation}, 15\penalty0 (6):\penalty0 1373--1396, 2003.
\newblock ISSN 0899-7667.

\bibitem[Busch et~al.(2023)Busch, Huang, Benz, Wallenstein, Lajoie, Wolf, Krishnaswamy, and {Turk-Browne}]{busch2023Multiview}
E.~L. Busch, J.~Huang, A.~Benz, T.~Wallenstein, G.~Lajoie, G.~Wolf, S.~Krishnaswamy, and N.~B. {Turk-Browne}.
\newblock Multi-view manifold learning of human brain-state trajectories.
\newblock \emph{Nature Computational Science}, 3\penalty0 (3):\penalty0 240--253, Mar. 2023.
\newblock ISSN 2662-8457.
\newblock \doi{10.1038/s43588-023-00419-0}.

\bibitem[Cannoodt(2018)]{Cannoodt2018princurve}
R.~Cannoodt.
\newblock Princurve 2.0: {{Fit}} a {{Principal Curve}} in {{Arbitrary Dimension}}.
\newblock CRAN, June 2018.

\bibitem[Crainiceanu et~al.(2016)Crainiceanu, Sweeney, Eloyan, and Shinohara]{crainiceanu2016tutorial}
C.~Crainiceanu, E.~M. Sweeney, A.~Eloyan, and R.~T. Shinohara.
\newblock A tutorial for multisequence clinical structural brain {{MRI}}.
\newblock In \emph{Handbook of Neuroimaging Data Analysis}, pages 109--133. CRC Press/Taylor \& Francis Group, Boca Raton, 2016.

\bibitem[Ding and Ma(2023)]{ding2023learning}
X.~Ding and R.~Ma.
\newblock Learning low-dimensional nonlinear structures from high-dimensional noisy data: {{An}} integral operator approach.
\newblock \emph{Ann. Statist.}, 51\penalty0 (4):\penalty0 1744--1769, Aug. 2023.
\newblock ISSN 0090-5364, 2168-8966.

\bibitem[Fefferman et~al.(2016)Fefferman, Mitter, and Narayanan]{fefferman2016testing}
C.~Fefferman, S.~Mitter, and H.~Narayanan.
\newblock Testing the manifold hypothesis.
\newblock \emph{Journal of the American Mathematical Society}, 29\penalty0 (4):\penalty0 983--1049, Oct. 2016.
\newblock ISSN 0894-0347, 1088-6834.
\newblock \doi{10.1090/jams/852}.

\bibitem[Gao et~al.(2023)Gao, Tao, Jaquier, and Asfour]{gao2023k}
J.~Gao, Z.~Tao, N.~Jaquier, and T.~Asfour.
\newblock K-{{VIL}}: {{Keypoints-Based Visual Imitation Learning}}.
\newblock \emph{IEEE Transactions on Robotics}, 39\penalty0 (5):\penalty0 3888--3908, Oct. 2023.
\newblock ISSN 1941-0468.

\bibitem[Green and Silverman(1994)]{greenSilverman1994}
P.~J. Green and B.~Silverman.
\newblock \emph{Nonparametric {{Regression}} and {{Generalized Linear Models}}: {{A Roughness Penalty Approach}}}.
\newblock Number~58 in Monographs on {{Statistics}} and {{Applied Probability}}. Springer, 1994.

\bibitem[Greven et~al.(2011)Greven, Crainiceanu, Caffo, and Reich]{greven2011longitudinal}
S.~Greven, C.~Crainiceanu, B.~Caffo, and D.~Reich.
\newblock Longitudinal {{Functional Principal Component Analysis}}.
\newblock In F.~Ferraty, editor, \emph{Recent {{Advances}} in {{Functional Data Analysis}} and {{Related Topics}}}, pages 149--154, Heidelberg, 2011. Physica-Verlag HD.
\newblock ISBN 978-3-7908-2736-1.

\bibitem[Hastie and Stuetzle(1989)]{hastiePrincipalCurves1989}
T.~Hastie and W.~Stuetzle.
\newblock Principal {{Curves}}.
\newblock \emph{J. Amer. Stat. Assoc.}, 84\penalty0 (406):\penalty0 502--516, June 1989.
\newblock ISSN 0162-1459, 1537-274X.

\bibitem[Hatcher(2002)]{hatcher2002algebraic}
A.~Hatcher.
\newblock \emph{Algebraic {{Topology}}}.
\newblock Cambridge University Press, 2002.

\bibitem[Knopman et~al.(2021)Knopman, Amieva, Petersen, Ch{\'e}telat, Holtzman, Hyman, Nixon, and Jones]{knopmanAlzheimerDisease2021}
D.~S. Knopman, H.~Amieva, R.~C. Petersen, G.~Ch{\'e}telat, D.~M. Holtzman, B.~T. Hyman, R.~A. Nixon, and D.~T. Jones.
\newblock Alzheimer disease.
\newblock \emph{Nature Reviews Disease Primers}, 7\penalty0 (1):\penalty0 1--21, May 2021.
\newblock ISSN 2056-676X.

\bibitem[Li et~al.(2023)Li, Tang, and Banerjee]{li2023inference}
D.~Li, W.~Tang, and S.~Banerjee.
\newblock Inference for {{Gaussian Processes}} with {{Matern Covariogram}} on {{Compact Riemannian Manifolds}}.
\newblock \emph{J. Mach. Learn. Res.}, 24\penalty0 (101):\penalty0 1--26, 2023.
\newblock ISSN 1533-7928.

\bibitem[Louis et~al.(2019)Louis, Couronn{\'e}, Koval, Charlier, and Durrleman]{louisRiemannianGeometryLearning2019}
M.~Louis, R.~Couronn{\'e}, I.~Koval, B.~Charlier, and S.~Durrleman.
\newblock Riemannian {{Geometry Learning}} for {{Disease Progression Modelling}}.
\newblock In A.~C.~S. Chung, J.~C. Gee, P.~A. Yushkevich, and S.~Bao, editors, \emph{Information {{Processing}} in {{Medical Imaging}}}, Lecture {{Notes}} in {{Computer Science}}, pages 542--553, Cham, 2019. Springer International Publishing.
\newblock ISBN 978-3-030-20351-1.
\newblock \doi{10.1007/978-3-030-20351-1_42}.

\bibitem[Meng and Eloyan(2021)]{mengPrincipalManifoldEstimation2021}
K.~Meng and A.~Eloyan.
\newblock Principal manifold estimation via model complexity selection.
\newblock \emph{J. R. Stat. Soc. Series B: (Statistical Methodology)}, 83\penalty0 (2):\penalty0 369--394, 2021.
\newblock ISSN 1467-9868.

\bibitem[Mulder et~al.(2014)Mulder, {de Jong}, Knol, {van Schijndel}, Cover, Visser, Barkhof, and Vrenken]{mulderHippocampalVolumeChange2014}
E.~R. Mulder, R.~A. {de Jong}, D.~L. Knol, R.~A. {van Schijndel}, K.~S. Cover, P.~J. Visser, F.~Barkhof, and H.~Vrenken.
\newblock Hippocampal volume change measurement: {{Quantitative}} assessment of the reproducibility of expert manual outlining and the automated methods {{FreeSurfer}} and {{FIRST}}.
\newblock \emph{NeuroImage}, 92:\penalty0 169--181, May 2014.
\newblock ISSN 1053-8119.
\newblock \doi{10.1016/j.neuroimage.2014.01.058}.

\bibitem[Muschelli et~al.(2015)Muschelli, Sweeney, Lindquist, and Crainiceanu]{muschelliFslrConnectingFSL2015}
J.~Muschelli, E.~Sweeney, M.~Lindquist, and C.~Crainiceanu.
\newblock Fslr: {{Connecting}} the {{FSL Software}} with {{R}}.
\newblock \emph{The R journal}, 7\penalty0 (1):\penalty0 163--175, June 2015.
\newblock ISSN 2073-4859.

\bibitem[Patenaude et~al.(2011)Patenaude, Smith, Kennedy, and Jenkinson]{patenaudeBayesianModelShape2011}
B.~Patenaude, S.~M. Smith, D.~N. Kennedy, and M.~Jenkinson.
\newblock A {{Bayesian}} model of shape and appearance for subcortical brain segmentation.
\newblock \emph{NeuroImage}, 56\penalty0 (3):\penalty0 907--922, June 2011.
\newblock ISSN 1053-8119.
\newblock \doi{10.1016/j.neuroimage.2011.02.046}.

\bibitem[{R Core Team}(2023)]{rSoftware2023}
{R Core Team}.
\newblock R: {{A Language}} and {{Environment}} for {{Statistical Computing}}.
\newblock R Foundation for Statistical Computing, 2023.

\bibitem[Roweis and Saul(2000)]{roweisNonlinearDimensionalityReduction2000}
S.~T. Roweis and L.~K. Saul.
\newblock Nonlinear {{Dimensionality Reduction}} by {{Locally Linear Embedding}}.
\newblock \emph{Science}, 290\penalty0 (5500):\penalty0 2323--2326, Dec. 2000.
\newblock \doi{10.1126/science.290.5500.2323}.

\bibitem[Smola et~al.(2001)Smola, Mika, Sch{\"o}lkopf, and Williamson]{smolaRegularizedPrincipalManifolds2001}
A.~J. Smola, S.~Mika, B.~Sch{\"o}lkopf, and R.~C. Williamson.
\newblock Regularized {{Principal Manifolds}}.
\newblock \emph{J. Mach. Learn. Res.}, 1\penalty0 (Jun):\penalty0 179--209, 2001.
\newblock ISSN ISSN 1533-7928.

\bibitem[Tenenbaum et~al.(2000)Tenenbaum, de~Silva, and Langford]{tenenbaumGlobalGeometricFramework2000}
J.~B. Tenenbaum, V.~de~Silva, and J.~C. Langford.
\newblock A {{Global Geometric Framework}} for {{Nonlinear Dimensionality Reduction}}.
\newblock \emph{Science}, 290\penalty0 (5500):\penalty0 2319--2323, Dec. 2000.

\bibitem[Wahba(1990)]{wahba1990}
G.~Wahba.
\newblock \emph{Spline models for observational data}.
\newblock SIAM, 1990.

\bibitem[Wolz et~al.(2010)Wolz, Aljabar, Hajnal, and Rueckert]{wolzManifoldLearningBiomarker2010}
R.~Wolz, P.~Aljabar, J.~V. Hajnal, and D.~Rueckert.
\newblock Manifold {{Learning}} for {{Biomarker Discovery}} in {{MR Imaging}}.
\newblock In F.~Wang, P.~Yan, K.~Suzuki, and D.~Shen, editors, \emph{Machine {{Learning}} in {{Medical Imaging}}}, Lecture {{Notes}} in {{Computer Science}}, pages 116--123, Berlin, Heidelberg, 2010. Springer.
\newblock ISBN 978-3-642-15948-0.
\newblock \doi{10.1007/978-3-642-15948-0_15}.

\bibitem[Yue et~al.(2016)Yue, Zipunnikov, Bazin, Pham, Reich, Crainiceanu, and Caffo]{yueParameterizationWhiteMatter2016}
C.~Yue, V.~Zipunnikov, P.-L. Bazin, D.~Pham, D.~Reich, C.~Crainiceanu, and B.~Caffo.
\newblock Parameterization of {{White Matter Manifold-Like Structures Using Principal Surfaces}}.
\newblock \emph{J. Amer. Stat. Assoc.}, 111\penalty0 (515):\penalty0 1050--1060, July 2016.
\newblock ISSN 0162-1459.

\end{thebibliography}

\newpage
\appendix
\section*{\bf{Supplementary Material for ``Longitudinal Principal Manifold Estimation"}}

\renewcommand\thefigure{S\arabic{figure}}
\renewcommand\thetable{S\arabic{table}}

\setcounter{figure}{0}
\setcounter{table}{0}

\begin{table}[h]
  \centering
  \begin{tabular}{|c c c c c|}
    \hline
    Case & Data & LPME & PME & PC/PS \\
    \hline
    1 & 0.223 (0.256) & {\bf 0.125 (0.161)} & 0.268 (0.850) & 0.189 (0.245) \\
    2 & 0.514 (0.408) & {\bf 0.384 (0.648)} & 0.843 (1.93) & 0.600 (0.296) \\
    3 & 0.446 (0.445) & {\bf 0.401 (0.446)} & 0.507 (0.594) & 0.412 (0.423) \\
    4 & 30.7 (88.1) & {\bf 27.7 (260)} & 30.7 (88.2) & 30.6 (88.1) \\
    5 & 0.980 (0.771) & {\bf 0.791 (0.845)} & 1.04 (1.07) & 0.934 (0.713) \\
    6 & 1.43 (6.04) & 1.21 (5.66) & 1.47 (6.06) & {\bf 1.01 (2.11)} \\
    7 & 0.580 (0.839) & 4.07 (3.30) & 7.37 (1.14) & {\bf 1.95 (0.800)} \\
    8 & 0.226 (0.275) & {\bf 0.136 (0.169)} & 0.242 (0.311) & 0.274 (0.243) \\
    \hline
  \end{tabular}
  \caption{MSD comparison to true values, Mean (SD). For each case, the lowest algorithm-specific mean (SD) are highlighted in bold. }
  \label{table:simulation_results_mean}
\end{table}

\begin{figure}

  \begin{subfigure}{\textwidth}
    \label{fig:adni_lhipp_result}
    \begin{subfigure}{\textwidth}
      \centering
      \includegraphics[height=3cm]{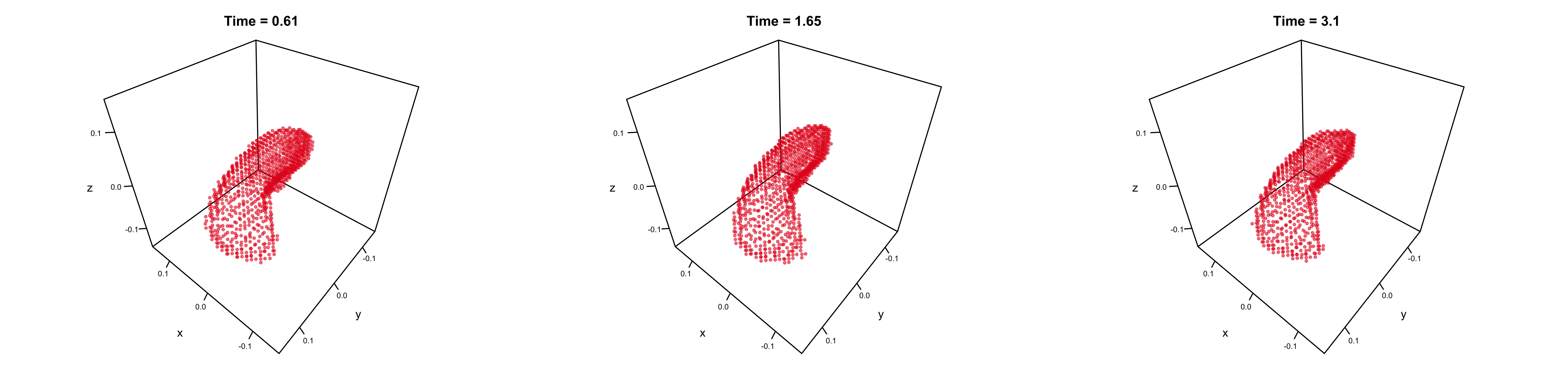}
      \caption{Data}
    \end{subfigure}
    \begin{subfigure}{\textwidth}
      \centering
      \includegraphics[height=3cm]{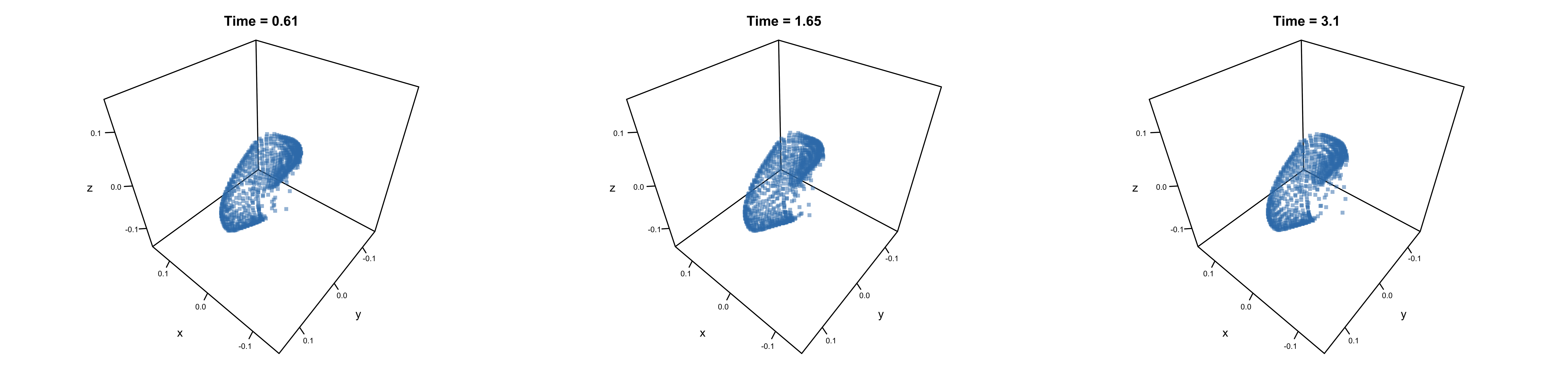}
      \caption{LPME}
    \end{subfigure}
    \begin{subfigure}{\textwidth}
      \centering
      \includegraphics[height=3cm]{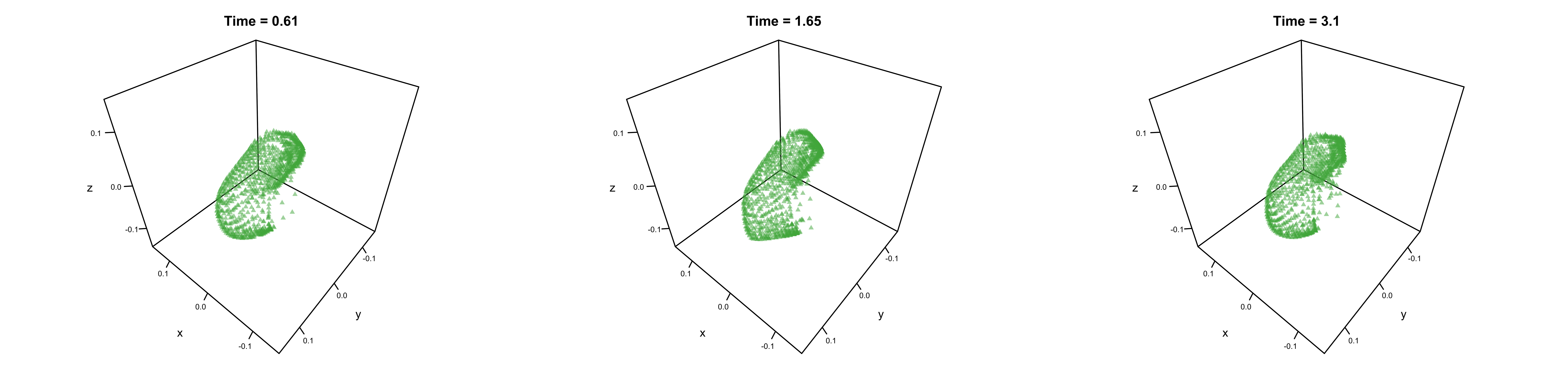}
      \caption{PME}
    \end{subfigure}
  \end{subfigure}

  \begin{subfigure}{\textwidth}
    \label{fig:adni_lthal_result}
    \begin{subfigure}{\textwidth}
      \centering
      \includegraphics[height=3cm]{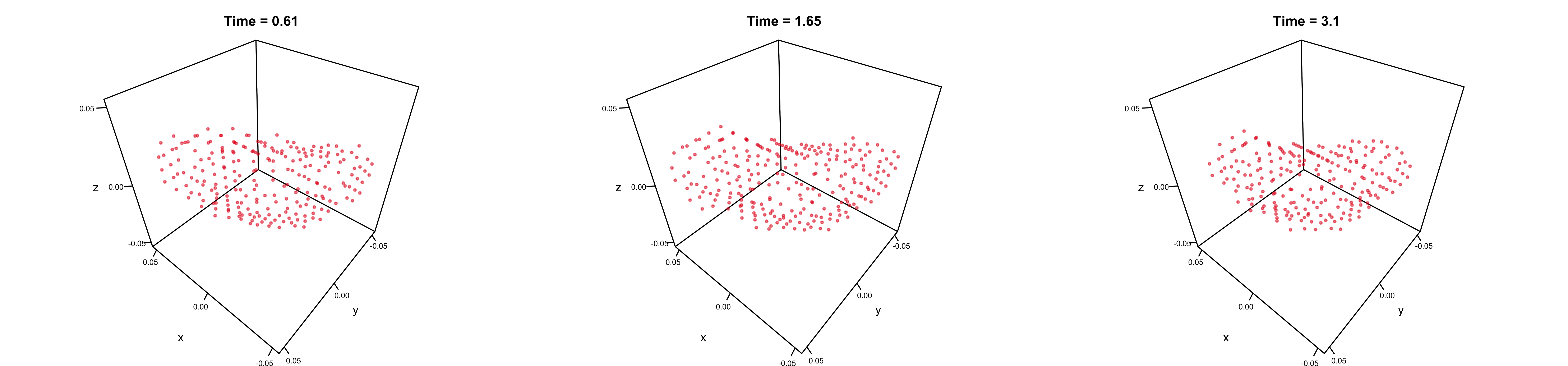}
      \caption{Data}
    \end{subfigure}
    \begin{subfigure}{\textwidth}
      \centering
      \includegraphics[height=3cm]{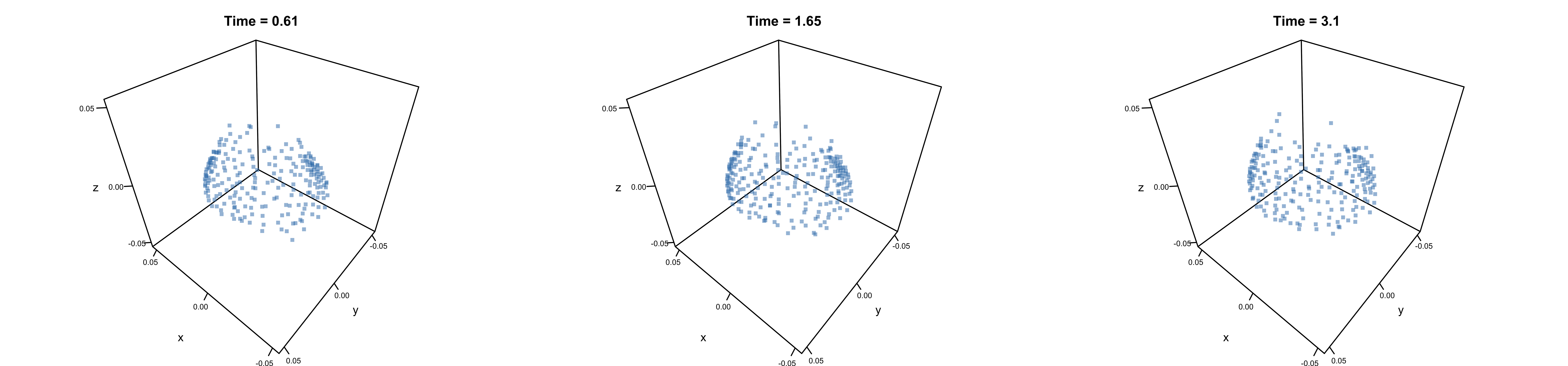}
      \caption{LPME}
    \end{subfigure}
    \begin{subfigure}{\textwidth}
      \centering
      \includegraphics[height=3cm]{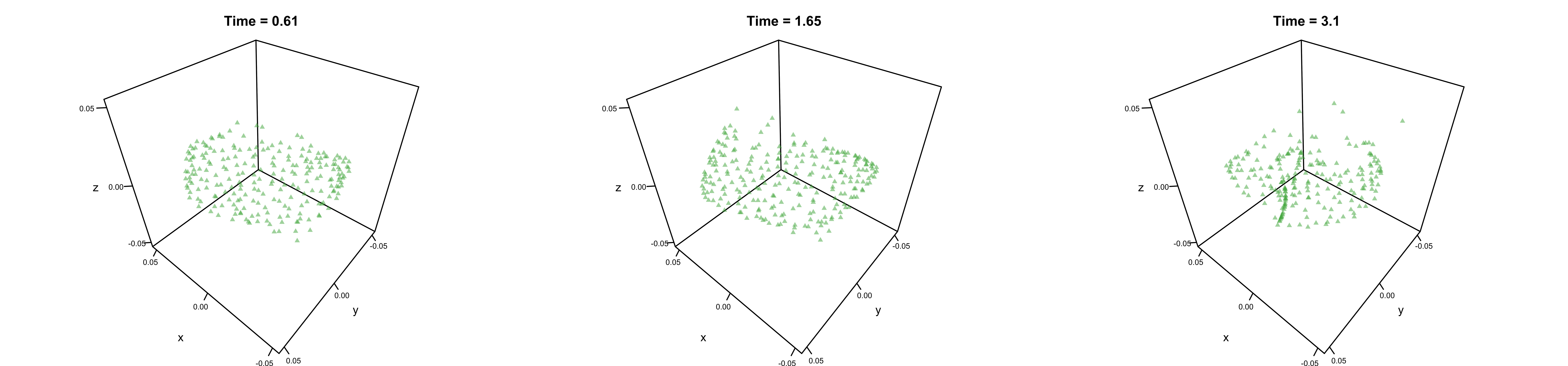}
      \caption{PME}
    \end{subfigure}
  \end{subfigure}

  \caption{Left Hippocampus (panels a-c), and Left Thalamus (panels d-f), Cognitive Healthy ADNI Participant. The raw surface data, displayed in red, show slight changes in orientation that are absent in the LPME estimates, shown in blue. The estimates obtained by the PME and LPME algorithms appear unable to accurately capture the true shape of the hippocampus at points with high levels of curvature.}
  \label{fig:adni_result}
\end{figure}

\begin{figure}
    \centering
    \subfloat[\centering Left Hippocampus Volume Estimates]{
      \includegraphics[height=8cm]{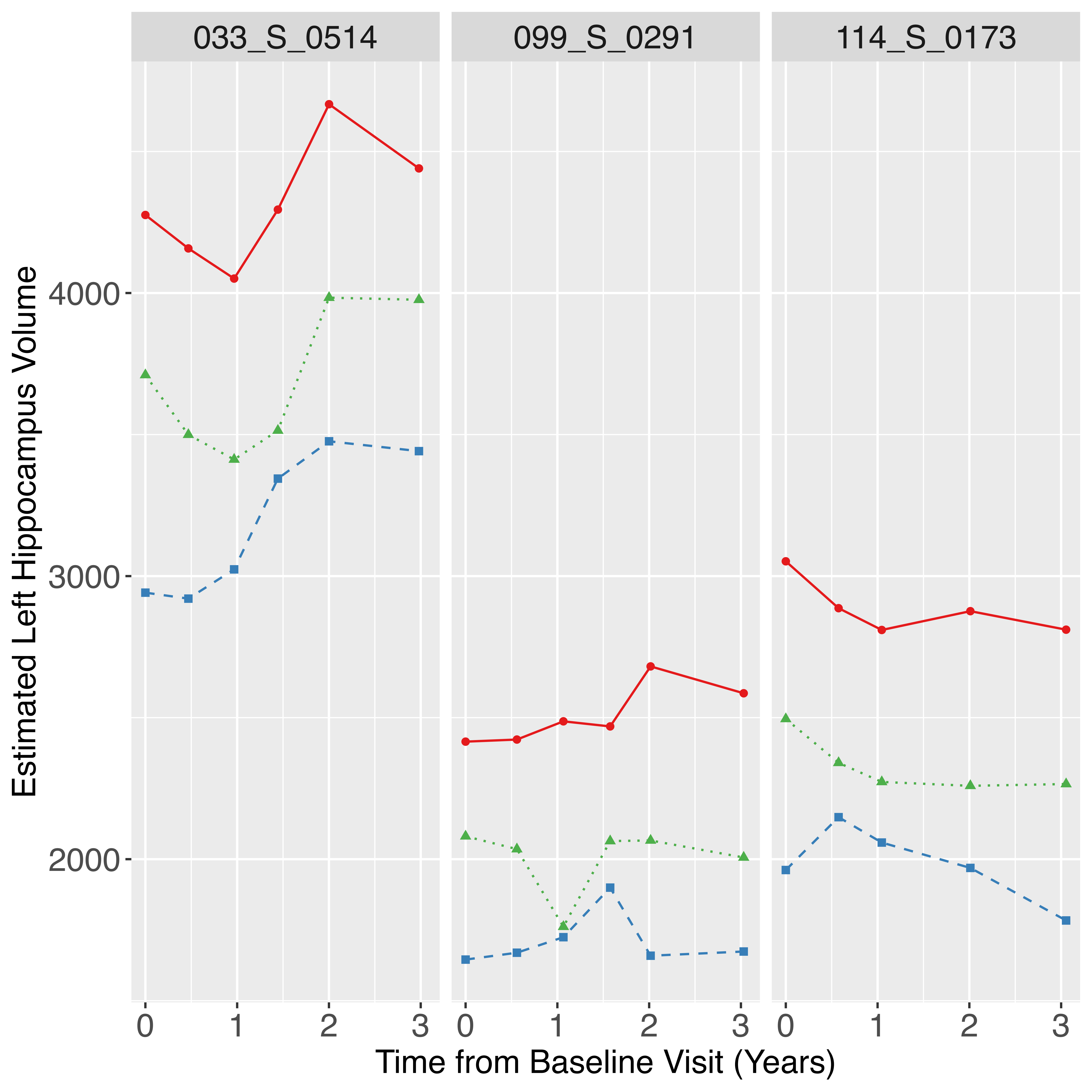}
      \label{fig:lhipp_volume_comparison}
    }
    \vfill
    \subfloat[\centering Left Thalamus Volume Estimates]{
      \includegraphics[height=8cm]{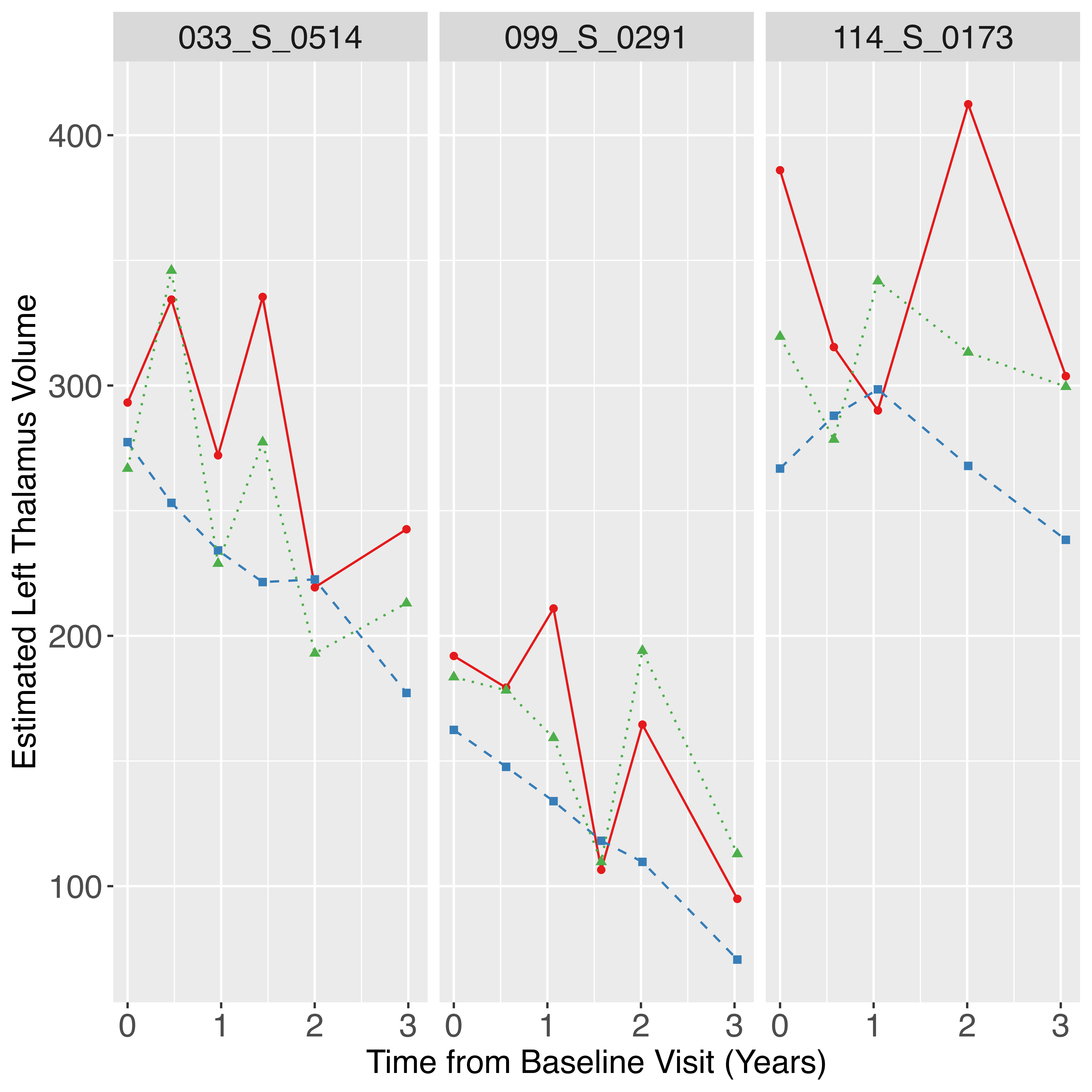}
      \label{fig:lthal_volume_comparison}
    }
    \caption{Left hippocampus and left thalamus volume estimates for three ADNI participants. Volume estimates are obtained by summing the volumes of voxels contained within the structure as described by the segmented data (circles, solid line), LPME (squares, dashed line), and PME (triangles, dotted line). When PME and LPME are applied to the irregular shape of the hippocampus, there is a clear ordering of the volume estimates. Gaps in the estimated surface induce underestimates of the volume compared to the volume values estimated from the data. When applied to the thalamus, the PME-based volume estimates demonstrate similar time point-to-time point changes in the volumes estimated from the segmented data, reflecting the close fit of the PME algorithm to the regularly-shaped thalamus data. The LPME estimates appear to successfully smooth over regions with large variations in subsequent volume measures from the data and PME estimates, as seen for participant 033\_S\_0514.}
\end{figure}

\end{document}